\documentclass[11pt]{article}
\usepackage{cite}
\usepackage{amsmath,amsfonts,amssymb}%,calrsfs}
\usepackage[small,bf,hang]{caption}
\usepackage{slashed}
\usepackage{color}

\usepackage{tikz-cd}

\usepackage{extarrows}
\usepackage{hyperref}

\def\hybrid{
        \topmargin -20pt
        \oddsidemargin 0pt
        \headheight 0pt \headsep 0pt
        \textwidth 6.25in % A4 paper
        \textheight 9.5in % A4 paper
        \marginparwidth .875in
        \parskip 5pt plus 1pt \jot = 1.5ex}

\hybrid

 \csname
@addtoreset\endcsname{equation}{section}

\newcommand{\p}{\partial}

\def\moth{\mathsurround=0pt}
\newdimen\zo \zo=0pt

\def\tick{\leaders\hrule height 0.5ex depth 0pt \hskip 0.5pt}
\def\upboxfill{$\moth \setbox\zo\hbox{\tick}%
  \hskip 3pt\hbox to 0pt{$\tick$\hss}\hrulefill \hbox to 7.5pt{$\tick$\hss}$}

\def\dtick{\leaders\hrule height .34pt depth 0.5ex \hskip 0.5pt}
\def\downboxfill{$\moth \setbox\zo\hbox{\dtick}%
  \hskip 2pt\hbox to 0pt{$\dtick$\hss}\hrulefill \hbox to 2pt{$\dtick$\hss}$}

\def\id{{\mathbb I}}

\def\bec{\begin{center}}
\def\ec{\end{center}}

\def\d{\delta} 

\def\be{\begin{equation}}
\def\ee{\end{equation}}
\def\bea{\begin{eqnarray}}
\def\eea{\end{eqnarray}}
\def\ba{\begin{array}}
\def\ea{\end{array}}

\begin{document}

\begin{titlepage}
\rightline{}
%\rightline\today
\rightline{December  2020}
\rightline{HU-EP-20/43-RTG}  
\begin{center}
\vskip 1.5cm
 {\Large \bf{ Gauge Invariant Perturbation Theory via Homotopy Transfer}}
\vskip 1.7cm

{\large\bf {Christoph Chiaffrino, Olaf Hohm and Allison F.~Pinto}}
\vskip 1.6cm

{\it  Institute for Physics, Humboldt University Berlin,\\
 Zum Gro\ss en Windkanal 6, D-12489 Berlin, Germany}\\
\vskip .1cm

\vskip .2cm

ohohm@physik.hu-berlin.de, chiaffrc@hu-berlin.de, apinto@physik.hu-berlin.de

\end{center}

\bigskip\bigskip
\begin{center} 
\textbf{Abstract}

\end{center} 
\begin{quote}

We show that the perturbative expansion of general gauge theories can be 
expressed in terms of gauge invariant variables to all orders in perturbations. 
In this we generalize techniques developed in gauge invariant cosmological 
perturbation theory, using Bardeen variables, by interpreting the passing over to 
gauge invariant fields as a homotopy transfer of the strongly homotopy Lie algebras 
encoding the gauge theory. This is illustrated for Yang-Mills theory, gravity on flat and cosmological backgrounds 
and for the massless sector of closed string theory. The perturbation lemma yields an algorithmic procedure to determine 
the higher corrections of the gauge invariant variables and the action in terms of these.

\end{quote} 
\vfill
\setcounter{footnote}{0}
\end{titlepage}

\tableofcontents

%\hfill \today

\section{Introduction}

One of the curious facts of modern physics is that its  fundamental theories, such as the  standard model of particle physics and Einstein's theory of general relativity, 
are gauge theories. The gauge symmetries defining these theories have a two-sided character: on the one hand, they uniquely determine the interactions 
and are hence a powerful tool in controlling various features of gauge field theories; on the other, they represent a mere redundancy of the formulation, 
which complicates the physical interpretation because  only \textit{gauge invariant} quantities may be observables.  
Logically, there are two ways of eliminating  this redundancy: one may impose conditions on the fundamental fields as gauge fixing conditions 
(making sure that these conditions can always be realized by legal gauge transformations) or one may aim to rewrite  the theory in terms of gauge invariant 
quantities. 
In this paper we will give an effective procedure in perturbation theory to construct gauge invariant variables to any order in fields
and to write the action in terms of these field variables, generalizing the methods of cosmological perturbation theory \cite{Bardeen:1980kt,Mukhanov:1990me,Kodama:1985bj}. 
While philosophically this seems to be quite different from simply fixing a gauge, 
our results also establish an operational equivalence with gauge fixing. 
In fact, when done properly, gauge fixing is just equivalent to introducing gauge invariant variables, as the latter formulation may always 
be reconstructed from the former. This will be explained and illustrated in more detail in the main text and the 
conclusion section.

Let us begin with some observations on the two ways of eliminating a gauge redundancy. 
The use of gauge fixing conditions is the standard text book approach in (quantum) gauge theory, but while conceptually straightforward 
it appears  to have 
certain  drawbacks. First, there are numerous possible gauge fixing conditions, making it somewhat of an art to pick  a convenient one  
(and leaving one with the agonizing, but presumably irrational, worry that the result in a different gauge may be different).
Second, even if a gauge fixing condition has been chosen, there are typically residual gauge transformations  and hence remaining redundancies that 
have to be dealt with. 
Most importantly, since observables have to be gauge invariant, the physical interpretation of results obtained  in a particular gauge may 
be unclear. These issues seem to have been particularly prominent  in cosmological  perturbation theory (see the classic review \cite{Mukhanov:1990me} by Mukhanov et.~al.~for a 
survey of the issues that arose historically). Maldacena, for instance, in his seminal computation of 
three-point functions  and non-Gaussianities, does the job twice,  in  two different gauges, just  to make sure that  the result is the same \cite{Maldacena:2002vr}. 
Clearly, it would be desirable to have a procedure that does not require  arbitrary gauge choices.

In first order cosmological perturbation theory, corresponding to the quadratic action around an FLRW background 
(as needed for the computation of two-point functions), there is a procedure introduced by Bardeen to formulate the 
theory directly in terms of gauge invariant field variables \cite{Bardeen:1980kt}, thereby removing the need to pick a gauge fixing condition. 
These techniques are based on a space/time split 
of fields and a decomposition  into irreducible components, such as scalar, vector or higher tensor modes, 
and so do not preserve manifest Lorentz invariance. It is common lore that gauge redundancies are only needed in order to 
make Lorentz invariance and locality manifest. The results of this paper provide an explicit procedure to eliminate any gauge redundancy, and in the process we indeed   
abandon manifest Lorentz invariance and introduce a certain non-locality (but only in space, not in time). 
While a loss of manifest Lorentz invariance may be too much to ask for in particle physics applications, in cosmology the situation is different:
there is a preferred (cosmic) time, so that  the space/time split needed for the definition of  Bardeen variables respects  the 
symmetries  of the FLRW backgrounds, hence making these variables ideally suited for cosmological perturbation theory.

Given this situation it seems desirable to extend gauge invariant cosmological perturbation theory beyond first order, 
as would be needed, for instance, in order to  compute the three-point functions of \cite{Maldacena:2002vr} in a manifestly  gauge invariant manner. 
General aspects of non-linear perturbation theory can be found  in \cite{Mukhanov:1996ak,Abramo:1997hu,Malik:2008im}, 
and in our approach we will recover some of their results. 
Specific results for second order perturbation theory  
have been 
obtained in \cite{Nakamura:2004rm,Giesel:2007wi,Giesel:2007wk,Weenink:2010rr,Prokopec:2012ug,Yoo:2014sfa,Domenech:2017ems,Chang:2020tji,Chang:2020iji}. 
However, a cursory scan of these references shows that the computations become quickly unwieldy and nontransparent. 
In this paper we will completely systematize gauge invariant perturbation theory to any order in perturbations by applying a powerful mathematical result known 
as homotopy transfer theorem or perturbation lemma. This is based on a formulation of gauge theories  in terms of strongly homotopy Lie or $L_{\infty}$ 
algebras \cite{Zwiebach:1992ie,Lada:1994mn}, 
which are generalizations of Lie algebras by  differentials  and higher brackets, see \cite{Lada:1992wc,Munster:2011ij} for self-contained reviews. 
The $L_{\infty}$ algebras encode the gauge transformations, field equations, 
Bianchi identities, etc., of the field theory. (See   \cite{Hohm:2017pnh} for the general dictionary and the introduction of \cite{Arvanitakis:2020rrk} for a concise 
summary. The $L_{\infty}$ formulation is also closely related to the classical limit of the Batalin-Vilkovisky (BV) formulation of field theory \cite{Batalin:1981jr,Batalin:1984jr}; 
more precisely, the tangent space to the BV manifold is the vector space of the $L_{\infty}$ algebra, 
and the $L_{\infty}$ brackets are recovered as the Taylor coefficients of the BV differential, which acts on the dual space of functions 
\cite{Alexandrov:1995kv}.)  
Homotopy transfer refers to a map from an $L_{\infty}$ algebra to another, generally `smaller' $L_{\infty}$ algebra that is homotopy equivalent 
(see \cite{vallette2014algebra,Arvanitakis:2020rrk} for pedagogical introductions and \cite{Doubek:2017naz,Jurco:2019yfd,Erbin:2020eyc,Koyama:2020qfb} for 
recent applications). 
We will show that passing over to gauge invariant Bardeen-type variables can be interpreted as homotopy transfer. The perturbation lemma then provides an 
effective and algorithmic procedure to determine gauge invariant variables to any order in perturbations.

In the remainder of the introduction we explain  some of our results in more technical detail. Let us consider Yang-Mills theory 
with gauge potential $A_{\mu}$, which  takes values in the Lie algebra of the gauge group. One starts by performing a space/time split $A_{\mu}=(A_{0},A_{i})$ 
and then uses the Helmholtz theorem familiar from 3d vector calculus to decompose the 3-vector into a divergenceless vector and the gradient of a scalar, 
$A_{i}=\widehat{A}_{i}+\partial_i\psi$, where $\partial_i\widehat{A}^i=0$.  
At the free linearized level, where the gauge transformations read $\delta A_{\mu}=\partial_{\mu}\Lambda$, one infers that the scalar $\psi$ transforms as 
$\delta \psi=\Lambda$, and so defining 
 \be\label{DEFIntrogaugeinv}
  \bar{A}_{\mu}=(\bar{A}_0,\bar{A}_i) \equiv (A_0-\dot{\psi}, A_{i}-\partial_i\psi)\,
 \ee
one quickly sees that these are gauge invariant field variables. 
It is illuminating to rewrite this as 
 \be\label{identitydecompforA}
  A_{\mu} = \bar{A}_{\mu}+\partial_{\mu}s(A)\;, 
 \ee
where $s(A)\equiv \psi$ is a  function of $A$ (that is non-local in space by picking out the scalar component of a vector-scalar decomposition). 
This function will be interpreted as the homotopy map that is needed for the application of the perturbation lemma. 
With the definition (\ref{DEFIntrogaugeinv}) this relation is just an identity, but it is a useful one because it takes a general unconstrained gauge field 
and decomposes it into a gauge invariant part plus a pure gauge part, with the `effective' gauge parameter being a (non-local) function of the field $A$ itself.  
Then replacing in the quadratic action every $A_{\mu}$ by the right-hand side of (\ref{identitydecompforA}) 
it follows by (linearized) gauge invariance that the $s(A)$ terms drop out, leaving the same action, but now in terms of $\bar{A}_{\mu}$, 
where we are allowed to use $\partial_i\bar{A}^i=0$, which follows from the definition of $\bar{A}_{\mu}$. 
We thus infer  that passing over to gauge invariant variables is 
operationally equivalent to imposing Coulomb gauge $\partial_iA^i=0$ and working out the action in this gauge. 
As we will see, analogous remarks apply to gravity on flat and cosmological backgrounds. While in the case of Maxwell's theory and even gravity 
linearized about flat space it is easy enough to rearrange the action in terms of gauge invariant variables, in the case of cosmological perturbation 
theory it simplifies the computation significantly already to first order  to use the above procedure for determining the action in terms of Bardeen variables 
(which, of course, gives the same result as the direct computation, which we also did).

As the main results of this paper we show, first, how the perturbation lemma can be used to define gauge invariant field variables to arbitrary order in 
perturbations and, second, that the action in terms of these gauge invariant variables is just obtained by replacing the original fields by the 
gauge invariant ones. We illustrate this for Yang-Mills theory by explicitly working out the gauge invariant field to next order. Denoting by $p$ the projection 
onto the part of the field that is gauge invariant to lowest order, i.e., $\bar{A}_{\mu}=p(A_{\mu})$, and by $\big[\cdot,\cdot\big]$
the Lie bracket of the Yang-Mills gauge algebra, the perturbation lemma yields for the gauge invariant variable to next order 
 \be\label{IntrINVREL}
  \widehat{A}_{\mu}=p\left(A_{\mu}+\tfrac{1}{2}\big[s(A),A_{\mu}+p(A_{\mu})\big]\right)\,. 
 \ee
Gauge \textit{invariance} (as opposed to gauge covariance)  may be verified by an explicit computation, which we also display in the main text. 
For the second result
we establish an identity, generalizing  (\ref{identitydecompforA}),  of the form 
 \be\label{NewIdentity}
  A_{\mu} \ = \ e^{\Delta_{\phi(A)}}\,\widehat{A}_{\mu}\;, 
 \ee 
where $\Delta$ is the operator defining  the infinitesimal Yang-Mills gauge transformation, i.e., 
$\Delta_{\Lambda}\widehat{A}_{\mu}=\partial_{\mu}\Lambda+\big[\widehat{A}_{\mu},\Lambda\big]$, 
and $\phi$ is a (non-local) function of $A$. 
This relation is again identically satisfied: we will give perturbative definitions for $\widehat{A}_{\mu}(A)$ and $\phi(A)$ so that 
writing out the right-hand side just reproduces $A_{\mu}$. 
Note that $\Delta_{\Lambda}\widehat{A}_{\mu}$ is \textit{not} actually the variation under infinitesimal gauge transformations, for this would vanish
due to $\widehat{A}_{\mu}$ being gauge invariant, which is the reason we denoted it by $\Delta_{\Lambda}$.
Nevertheless, since (\ref{NewIdentity}) takes the form of a finite gauge transformation it follows by gauge invariance that  in the non-linear action 
we may simply replace $A_{\mu}$ by the gauge invariant $\widehat{A}_{\mu}$. This result is so enticingly simple to be almost disappointing, for 
it means that, to all orders in perturbations, writing the action in terms of gauge invariant variables is operationally  equivalent to imposing a gauge fixing condition: 
Coulomb gauge in the present example, for $\partial_i\widehat{A}^{i}=0$ holds to all orders, or transverse gauge for the Bardeen 
variables of cosmological perturbation theory. (As we will outline in the conclusion  section, in principle one may also define gauge invariant variables that mimic 
other gauge conditions, although for cosmology the Bardeen variables seem to be preferred.)  
Nevertheless, our results also show  
how to reconstruct from a computation   in a particular gauge the fully gauge invariant result written in gauge invariant variables, 
thereby hopefully resolving some of the interpretational issues alluded to above. 
More generally, we hope that our interpretation of gauge invariant variables in terms of homotopy transfer and the corresponding availability of 
powerful mathematical techniques will help systematize the perturbative treatment  of various problems, 
particularly in gravitational physics.

The rest of this paper is organized as follows. In sec.~2 we illustrate gauge invariant variables  for the simple toy models
of abelian spin-1 and spin-2 gauge theories (Maxwell's theory and Fierz-Pauli theory) and explain how this can be formulated 
in terms of chain complexes and homotopy transfer. While the writing of  these theories  in terms of gauge invariant variables 
is completely  straightforward, we have not been able to find them in the literature before. 
Then, in sec.~3, we work out the example of double field theory on a flat background, which shows some new features 
by having an extended  chain complex,  which is necessary due to the gauge algebra not being a Lie algebra but a genuine 
$L_{\infty}$ algebra. The general theory of $L_{\infty}$ algebras and homotopy transfer is reviewed in sec.~4 in a hopefully self-contained manner. 
We then show how the homotopy transfer theorem allows for the construction of gauge invariant variables to all orders for arbitrary gauge theories
and illustrate this explicitly for the example of Yang-Mills theory. This is the core technical result of this paper. 
 In sec.~5 we illustrate the power of these techniques by determining the complete 
quadratic action for cosmological perturbation theory in terms of gauge invariant variables  by the 
simple procedure explained above. To the best of our knowledge, this general action 
 has not been given in the literature before. Finally, in sec.~6 we close with a brief discussion and outlook.

\section{Gauge invariant spin-1 and spin-2 theory}

In this section we illustrate the general approach for two simple theories: Maxwell's theory and gravity linearized about flat space. 
We explain the formulation of these theories in terms of chain complexes, introduce gauge invariant variables and establish the 
notion of homotopy transfer.

\subsection{Maxwell's theory}

We start with Maxwell's theory for a gauge potential $A_{\mu}$ with action functional 
 \be\label{MaxwellAction}
  S= -\frac{1}{4}\int {\rm d}^4x \,F^{\mu\nu}F_{\mu\nu}\;, 
 \ee
where $F_{\mu\nu}=\partial_{\mu}A_{\nu}-\partial_{\nu}A_{\mu}$ is the field strength. This theory is invariant 
under the abelian gauge transformations $\delta A_{\mu}=\partial_{\mu}\Lambda$. 
We aim to rewrite the theory in terms of gauge invariant variables. To this end we perform a space/time split of the gauge field, 
writing $A_{\mu}=(A_0,A_i)$, and employ the  Helmholtz theorem according to which 
any 3-vector $A_i$ can be decomposed into 
a divergenceless vector plus the gradient of a scalar: 
 \be\label{SVTvector}
  A_{i}=\widehat{A}_{i}+\partial_i\psi\,, \qquad \partial_i\widehat{A}^{i}=0\,. 
 \ee
Effectively, this decomposition exists whenever the spatial Laplacian $\Delta\equiv \partial^i\partial_i$ is invertible 
(say in terms of the familiar integral expression using the Green's function), for then we can set 
 \be\label{psiDEF}
  \psi=\Delta^{-1}(\partial_i A^{i})\,,
 \ee
as follows by taking the divergence of (\ref{SVTvector}). This is the (space) non-locality inherent in this approach. 
One may read off the gauge transformations of the vector $\widehat{A}_i$ and scalar $\psi$:
 \be
   \delta A_0 = \dot{\Lambda}\,, \qquad \delta \widehat{A}_{i}=0\,,\qquad \delta \psi=\Lambda\,. 
 \ee
Thus, $\widehat{A}_{i}$ is gauge invariant. Furthermore, we observe that the combination 
 \be\label{PhiDEfgaugeIVVV}
  \Phi \equiv A_0-\dot{\psi}\; 
 \ee
is gauge invariant. It is now straightforward to verify that the Maxwell equations $\partial_{\mu}F^{\mu\nu}=0$ can 
be written in the manifestly gauge invariant form 
 \be
 \begin{split}
  \Delta \Phi &=0\,, \\
  \Box\, \widehat{A}_{i} + \partial_i\dot{\Phi} &= 0\,. 
 \end{split} 
 \ee
Under the same assumption of invertibility of $\Delta$ the first equation implies $\Phi=0$, so that the second equation yields 
the equation $\Box\, \widehat{A}_{i}=0$ for the two independent degrees of freedom encoded in $\widehat{A}_{i}$. 
The Maxwell action (\ref{MaxwellAction}) can be written in the manifestly gauge invariant form 
 \be
  S=\frac{1}{2}\int {\rm d}^4x \big(\widehat{A}^{i}\,\Box\,\widehat{A}_{i}-\Phi\Delta \Phi\big)\,. 
 \ee

Next, we  briefly discuss  the inclusion of matter, in which case the non-dynamical $\Phi$ is determined by the charge density
and hence non-vanishing. Specifically,  we can couple a current $j^{\mu}=(\rho,\widehat{j}^{i}+\partial^i\sigma)$, with $\partial_i\widehat{j}^{i}=0$,  
for which the conservation law $\partial_{\mu}A^{\mu}=0$ 
implies $\dot{\rho}+\Delta \sigma=0$. Using this one may verify that adding  the matter coupling $A_{\mu}j^{\mu}$ to (\ref{MaxwellAction}) results in 
the manifestly gauge invariant action 
 \be
  S=\int {\rm d}^4x \Big(\frac{1}{2}\widehat{A}^{i}\,\Box\,\widehat{A}_{i}-\frac{1}{2} \Phi\Delta \Phi -\Phi\rho-\widehat{A}_{i}\widehat{j}^{i}\Big)\,. 
 \ee

\subsubsection*{Homotopy interpretation}

We now give an algebraic  interpretation of the passing over to gauge invariant variables in terms of homotopy transfer. 
We first explain  that the `data' of a free gauge theory, such as gauge transformations, equations of motion, and Bianchi identities, 
 can be encoded in a so-called chain complex. This is a sequence of vector spaces $X_i$ with maps (abstract differentials) $\partial_i:X_i\rightarrow X_{i-1}$ 
 that square to zero in the sense that $\partial_{i-1}\circ \partial_i=0$. 
(We note that this encodes the same information as the BV-BRST  differential, see, e.g., \cite{Fisch:1989rp,Barnich:2004cr}.) 
 For Maxwell's theory the chain complex reads 
  \be\label{chaimcomplexMaxwell}
\begin{array}{ccccccccccc} X_1 &\xlongrightarrow{\partial_1} &X_0 {}  &\xlongrightarrow{\partial_0}
&X_{-1} &
\xlongrightarrow{\partial_{-1}}& X_{-2}
\\[1.5ex]
\{\Lambda\}& &\{A_{\mu}\} & &\{E^{\mu}\}&
&\{f\}
\end{array}
\ee
where we have indicated in the second line the interpretation of each space: $X_1$ is the space of gauge parameters $\Lambda$, 
$X_0$ is the space of gauge fields $A_{\mu}$, $X_{-1}$ is the space of field equations $E^{\mu}$, and $X_{-2}$ a space of scalars 
encoding the Bianchi identity. The differentials are defined as 
 \be\label{MaxwellDifferentials}
 \begin{split}
  \partial_1(\Lambda)_{\mu} \ &= \ \partial_{\mu}\Lambda\,,\\
  \partial_0(A)_{\mu}\  &= \ \Box\, A_{\mu}-\partial_{\mu}(\partial^{\nu}A_{\nu})\,,\\
  \partial_{-1}(E) \ &= \ \partial_{\mu}E^{\mu}\;. 
 \end{split}
 \ee 
The gauge transformations and field equations are then encoded in $\delta A=\partial_1(\Lambda)$ and $\partial_0(A)=0$, respectively. 
The nilpotency of the differential amounts to $\partial_0\circ \partial_{1}=0$ and $\partial_{-1}\circ \partial_0=0$, 
which encode (linearized) gauge invariance and the Bianchi identity, respectively. 
(In the following we will often skip the subscript on $\partial$ and similar maps as the source and target spaces are  usually clear from the context.)

Next we define projection maps $p$ from this chain complex to the complex of gauge invariant variables:
 \be\label{DIAGRAM}
\begin{array}{ccccccccccc}X_1 &\xlongrightarrow{\partial} &X_0 {}  &\xlongrightarrow{\partial}
&X_{-1} &
\xlongrightarrow{\partial}& X_{-2}
\\[1.5ex]
\Big{\downarrow}{p}&&\Big{\downarrow}{p}&&\Big{\downarrow}{p}&&\Big{\downarrow}{p}
\\[1.5ex]
\{0\}&\xlongrightarrow{\bar{\partial}=0}&\bar{X}_{0} &\xlongrightarrow{\bar{\partial}}&\bar{X}_{-1}&
\xlongrightarrow{\bar{\partial}=0}&\{0\}
\end{array}
\ee
Note, in particular, that the space of gauge parameters is projected to zero, indicating that there are no gauge redundancies left, 
in agreement with the interpretation of $\bar{X}_0$ as the space of gauge invariant fields. For the fields the projector is defined by 
 \be\label{gaugeinvProjA}
  \bar{A}_{\mu}=p(A_{\mu})=(\bar{A}_0,\bar{A}_{i})\equiv (\Phi,\widehat{A}_{i})\,,
 \ee
and consists of the gauge invariant variables defined above. 
For the space $X_{-1}$ of field equations the projector is 
 \be\label{projjjEE}
  \bar{E}^{\mu}=p(E^{\mu})= E^{\mu}-(0,\partial^i\Delta^{-1}(\partial_{\mu}E^{\mu}))\,. 
 \ee
From this definition it follows 
 \be
  \partial_{\mu}\bar{E}^{\mu}\equiv 0\,, 
 \ee 
i.e., the projected space  $\bar{X}_{-1}$ consists of 4-vectors with zero divergence. All other projection maps in (\ref{DIAGRAM}) are trivial. 
We also have to introduce inclusion maps going in the other direction, $\iota: \bar{X}\rightarrow X$, that simply 
view a projected object as an element of the original space, i.e., $\iota(\bar{A}_{\mu})=\bar{A}_{\mu}$ and $\iota(\bar{E}^{\mu})=\bar{E}^{\mu}$. 

We are now ready to explain the homotopy relation that is the basis for all our subsequent applications. 
We first note that, by definition, the composition $p\circ \iota$ equals the identity on $\bar{X}$, but 
in general the composition $\iota\circ p$ going the other direction does not equal the identity on $X$ since 
the projection of course shrinks the space. However, in the case at hand $\iota\circ p$ is equal to the identity \textit{up to homotopy}, 
which means that there are degree $+1$ maps $s_i: X_i\rightarrow X_{i+1}$ (going in the opposite direction of the differential $\partial$) 
so that 
 \be\label{homotopyRelation}
  (\iota \circ p)_i = {\rm id}_{X_i}-\partial_{i+1}\circ s_i-s_{i-1}\circ \partial_i\;. 
 \ee
The significance of this relation, to be explained in sec.~4, is that an $L_{\infty}$ structure on $X$ (encoding non-linear gauge symmetries and dynamics) 
can then be transported to an $L_{\infty}$ structure on $\bar{X}$ (for instance encoding non-linear dynamics without any gauge symmetries left). 

Let us now define the maps $s_i: X_i\rightarrow X_{i+1}$ so that (\ref{homotopyRelation}) holds. 
Evaluating the homotopy relation on fields $A_{\mu}\in X_0$ one obtains 
 \be\label{homotopyonAAA}
  \iota p(A_{\mu}) = \bar{A}_{\mu}=A_{\mu}-\partial_{\mu}(s_0(A))\,,
 \ee
where we assumed 
 \be\label{HonEiszero}
  s_{-1}\equiv 0\,,
 \ee
whose consistency will be confirmed momentarily. Recalling (\ref{gaugeinvProjA}) and the definition (\ref{PhiDEfgaugeIVVV}) 
of gauge invariant variables we have 
 \be
  \bar{A}_{\mu}=(\Phi,\widehat{A}_{i}) = (A_0-\dot{\psi}, A_{i}-\partial_i\psi)=A_{\mu}-\partial_{\mu}\psi\,.
 \ee
Therefore, the homotopy relation (\ref{homotopyonAAA}), which we also write as 
 \be
  A_{\mu}=\bar{A}_{\mu}+\partial_{\mu}(s_0(A))\;, 
 \ee 
is obeyed for 
 \be\label{homotopyon0}
  s_0(A)=\psi =\Delta^{-1}(\partial_iA^i)\;, 
 \ee
where we recalled the definition (\ref{psiDEF}) of $\psi$ in terms of $A_{\mu}$. 
Next, we evaluate the homotopy relation on $E^{\mu}\in X_{-1}$, 
 \be
  \iota p(E^{\mu})=\bar{E}^{\mu} = E^{\mu}  -s^{\mu}(\partial\cdot E)\;, 
 \ee
where we used (\ref{HonEiszero}) and the definition (\ref{MaxwellDifferentials}) of the differential $\partial(E)\equiv \partial\cdot E\equiv \partial_{\mu} E^{\mu}$. 
Recalling the projection (\ref{projjjEE}) of $E^{\mu}$ one infers that the homotopy relation is obeyed upon setting for arbitrary $f\in X_{-2}$
 \be\label{Honchispace}
  s^{\mu}(f) = (0,\partial^i\Delta^{-1}f)\;. 
 \ee
Finally, having fixed the homotopy maps,  we have to verify the homotopy relation on $X_{1}$ and $X_{-2}$. 
For $\Lambda\in X_1$ we compute 
 \be
  \iota p(\Lambda) = 0 = \Lambda-s(\partial \Lambda)\,,
 \ee
which is obeyed thanks to $s(\partial \Lambda)=\Delta^{-1}\partial_i\partial^i\Lambda=\Lambda$, upon using (\ref{homotopyon0}). 
On $f\in X_{-2}$ one finds 
 \be
  \iota p(f)=0=f-\partial_{\mu}(s^{\mu}(f))\;, 
 \ee
which indeed is identically satisfied for (\ref{Honchispace}). 
Summarizing, we have shown that for homotopy maps $s_i$ defined by (\ref{HonEiszero}), (\ref{homotopyon0}) and (\ref{Honchispace}) 
the homotopy relation (\ref{homotopyRelation}) is satisfied. This proves that passing over to gauge invariant variables can be interpreted 
as homotopy transfer.

\subsection{Gravity on flat space} 

We now redo the above analysis for gravity linearized about flat Minkowski space, as a preparation for the technically more involved 
case of cosmological perturbation theory. 
 Expanding the spacetime metric 
as $g_{\mu\nu}=\eta_{\mu\nu}+h_{\mu\nu}$, with $\eta_{\mu\nu}$ the Minkowski metric, 
to second order in fluctuations, the Einstein-Hilbert action reduces to 
the Fierz-Pauli action
\begin{equation} \label{FPactionG}
S_{\rm FP} = - \frac{1}{2} \int {\rm d}^4x\, h^{\mu\nu} G_{\mu\nu} ( h) \,, 
\end{equation} 
where $G_{\mu\nu}$ denotes the linearized Einstein tensor 
 \be\label{linEinstein}
  G_{\mu\nu} = R_{\mu\nu}-\frac{1}{2}R\,\eta_{\mu\nu}\;, 
 \ee 
with the linearized Ricci tensor
\begin{equation}
   R_{\mu\nu} \equiv  -\frac{1}{2}\big(\Box\, h_{\mu\nu} -2\,\partial^{\rho}\partial_{(\mu} h_{\nu)\rho} + \partial_{\mu}\partial_{\nu}h\big)\;, 
  \end{equation}
and the linearized scalar curvature 
  \be
  R \equiv  \eta^{\mu\nu} R_{\mu\nu} = \partial^{\mu}\partial^{\nu}h_{\mu\nu}-\Box\, h \;. 
  \ee
 The Fierz-Pauli action is invariant under linearized diffeomorphisms 
  \be\label{linDIff}
   \delta h_{\mu\nu}=\partial_{\mu}\xi_{\nu}+\partial_{\nu}\xi_{\mu}\,, 
  \ee 
a fact that is also expressed in  the Bianchi identity $\partial^{\mu} G_{\mu\nu}\equiv 0$. 

As for Maxwell's theory, we will next rewrite linearized gravity in terms of gauge invariant variables, illustrating in a simpler 
setting the techniques of cosmological perturbation theory. In the same way that 
we decomposed a 3-vector into a divergence-free vector plus the gradient of a scalar one begins with a 
 so-called scalar-vector-tensor (SVT) decomposition. Here, a symmetric tensor is decomposed into a traceless and 
divergence-free tensor plus the trace part plus   the gradients of vectors and scalars. Specifically, we write 
\be\label{SVTdecompositionTensor}
 \begin{split}
  h_{00}&=-2\phi\,,\\ 
  h_{0i}&= {B}_{i}+\partial_iB\,,\\
  h_{ij} &= \widehat{h}_{ij}+  2 C\delta_{ij} + \partial_{i}{E}_{j}+\partial_{j}{E}_{i} +2(\partial_i\partial_jE-\tfrac{1}{3}\delta_{ij}\Delta E)\;, 
 \end{split}
\ee
subject to the constraints 
 \be\label{constraintsonBEandh}
  \partial^{i}\widehat{h}_{ij} = \widehat{h}^{i}{}_{i}= \partial^i{E}_{i}= \partial^{i}B_i= 0\,. 
 \ee
We next determine the gauge transformations of the components fields, where 
we subject the gauge parameter to the same decomposition $\xi_{\mu}=(\xi_0,\xi_i)$, where   
 \be\label{decompositionxi}
  \xi_{i}=\zeta_{i}+\partial_{i}\chi\,, \qquad \partial_i\zeta^i=0\,. 
 \ee 
One finds with (\ref{linDIff}) 
 \be\label{splitgaugestuff}
 \begin{split}
  \delta h_{00}&=2\,\dot{\xi}_0\,,\\
  \delta h_{0i} &= \dot{\zeta}_i+\partial_i(\dot{\chi}+\xi_0)\,,\\
  \delta h_{ij} &= \partial_i\zeta_j +\partial_j\zeta_i+2\,\partial_i\partial_j\chi\,. 
 \end{split} 
 \ee 
From the first equation in (\ref{SVTdecompositionTensor}) we then immediately infer 
 \be\label{deltaPhi}
  \delta \phi = -\dot{\xi}_0\,. 
 \ee 
Next, varying the second equation in (\ref{SVTdecompositionTensor}), 
 \be
  \delta h_{0i}= \delta {B}_{i}+\partial_i\delta B=\dot{\zeta}_i+\partial_i(\dot{\chi}+\xi_0)\;, 
 \ee
and comparing the divergence-free and gradient parts on both sides of the equation  we obtain 
 \be\label{deltaBBB}
   \delta {B}_{i} = \dot{\zeta}_{i}\,, \qquad 
   \delta B = \dot{\chi}+\xi_0\,. 
 \ee 
A similar analysis  
gives the following gauge transformations for the components of $h_{ij}$: 
 \be\label{deltahij}
  \begin{split}
   \delta {E}_{i}&=\zeta_i\,,\qquad \,\,
   \delta C = \frac{1}{3}\,\Delta\chi\,,\\
   \delta E&=\chi\,,\qquad  
   \delta \widehat{h}_{ij} = 0\,. 
  \end{split}
 \ee  
We learn that $\widehat{h}_{ij}$ is gauge invariant. In addition, it is easy to see with (\ref{deltaPhi}), (\ref{deltaBBB}) and (\ref{deltahij}) 
that we can build three more gauge invariant combinations: 
 \be\label{gaueinvgravitymodes}
  \begin{split}
   \Sigma_i \equiv \dot{{E}}_{i}-B_{i}\,, \qquad 
   \Psi \equiv - C + \frac{1}{3} \Delta E \,, \qquad 
   \Phi\equiv \phi+\dot{B}- \ddot{E}\,. 
  \end{split}
 \ee
It should be noted that these quantities are not unique in that any function  of gauge invariant objects is also gauge invariant. 
It is instructive to count the number of gauge invariant quantities (say in four dimensions):  $\widehat{h}_{ij}$ has $6-3-1=2$ components (subtracting the 
trace and divergence constraints from the six components of a symmetric 2-tensor in three-dimensional space), $\Sigma_i$ has $2$ components 
(since it is divergence-free), while $\Phi$ and $\Psi$ each have one degree of freedom, making a total of six degrees of freedom.  
This is what one should expect, since we started with the ten components of $h_{\mu\nu}$ which are subject to gauge redundancies 
with parameters $\xi_{\mu}$ having four components.

We can now write the vacuum Einstein equations in terms of the gauge invariant quantities  $\widehat{h}_{ij}, \Sigma_i, \Phi$ and $\Psi$. 
After some algebra one finds for the components of the Einstein tensor (\ref{linEinstein}):  
 \be\label{gaugeinvaGcomponents}
  \begin{split}
   G_{00} &=2 \Delta \Psi \,,\\
   G_{0i} &= \frac{1}{2}\Delta \Sigma_i +2\partial_i\dot{\Psi}\,,\\
   G_{ij} &= -\frac{1}{2}\Box\, \widehat{h}_{ij} -\partial_i\partial_j\big(\Phi- \Psi\big) 
   +\delta_{ij}\big(\Delta \Phi- \Delta \Psi
  +2\ddot{\Psi}\big) + \partial_{(i}\dot{\Sigma}_{j)}\,. 
  \end{split}
 \ee
It is also instructive to analyze the gauge invariant content of linearized gravity using these equations.  
From the equation $G_{00}=0$ we infer $\Delta \Psi=0$, 
which by the assumed invertibility of $\Delta$ tells us $\Psi=0$. Using this in the second equation $G_{0i}=0$ then tells us $\Delta \Sigma_i=0$ and hence 
that $\Sigma_i=0$. The final equation $G_{ij}=0$ then reduces to 
 \be
  -\frac{1}{2}\Box \,\widehat{h}_{ij}-\partial_i\partial_j \Phi +\delta_{ij} \Delta \Phi =0\,. 
 \ee
Taking the trace of this equation (recalling that $\widehat{h}_{ij}$ is traceless) we learn $\Delta \Phi=0$ and hence $\Phi=0$, so that 
the equations finally reduce to 
 \be\label{finalWAVE}
  \Box\, \widehat{h}_{ij} = 0\,. 
 \ee 
We thus find, as expected, that the gravitational field $h_{\mu\nu}$ carries the two propagating degrees of freedom encoded in the 
divergence-free and traceless $\widehat{h}_{ij}$.

Let us now analyze the action in terms of gauge invariant quantities. Inserting (\ref{gaugeinvaGcomponents}) into the Fierz-Pauli 
action (\ref{FPactionG}) it is straightforward to bring it into the manifestly gauge invariant form 
 \be
  {\cal L}_{\rm FP} = \frac{1}{4}\widehat{h}^{ij}\,\Box \,\widehat{h}_{ij} -\frac{1}{2}\Sigma^i \Delta \Sigma_i + (4\Phi-2\Psi)\Delta \Psi
  +6\Psi\ddot{\Psi}\,. 
 \ee
This action is not diagonal, but it can be diagonalized by a suitable (space non-local) field redefinition. 
It is convenient to do this computation right away in the more general setting where one includes matter sources via an energy-momentum tensor $T^{\mu\nu}$. 
The energy-momentum tensor  may also be subjected to an SVT decomposition: 
 \be
  \begin{split}
   T^{00}&=\rho\,, \\
   T^{0i} &= q^i+\partial^i q\,, \quad \partial_iq^i=0\,, \\
   T^{ij} &= \Pi^{ij}+\partial^i\Pi^j+\partial^j\Pi^i+\partial^i\partial^j \Pi -\frac{1}{3}\delta^{ij}\Delta \Pi +p\delta^{ij}\,, 
  \end{split}
 \ee 
where 
 \be
  \Pi^{i}{}_{i}=\partial_i\Pi^{ij}=\partial_i\Pi^{i}=0\,. 
 \ee 
The energy-momentum conservation law $\partial_{\mu}T^{\mu\nu}=0$ then implies 
 \be\label{energymomentum}
  \begin{split}
   \dot{\rho}+\Delta q &=0\,,\\
   \dot{q}^{j}+\Delta \Pi^{j} &= 0\,,\\
   p+\dot{q}+\frac{2}{3}\Delta \Pi &=0\,. 
  \end{split}
 \ee 
Using these one finds for the coupling to gravity:
 \be
  \frac{1}{2}h_{\mu\nu}T^{\mu\nu}=\frac{1}{2}\widehat{h}_{ij}\Pi^{ij} -\Sigma_i q^i-\Phi\rho+\frac{3}{2}\Psi\, p\,, 
 \ee  
and hence for the complete Lagrangian: 
 \be
 \begin{split}
  {\cal L}_{\rm FP, \, matter} = &\,\frac{1}{4}\widehat{h}^{ij}\,\Box \,\widehat{h}_{ij} -\frac{1}{2}\Sigma^i \Delta \Sigma_i 
  + (4\Phi-2\Psi)\Delta \Psi
  +6\Psi\ddot{\Psi} \\
  &+\frac{1}{2}\widehat{h}_{ij}\Pi^{ij} -\Sigma_i q^i-\Phi\rho-3\Psi\, p\,. 
 \end{split}
 \ee
We note that the components of the energy-momentum tensor $q$, $\Pi^i$, and $\Pi$ have dropped out, 
i.e., we lost four degrees of freedom, which matches the number  of conservation constraints.

Our next task is to diagonalize this action so that it expresses the physical content directly. 
We can perform the following field redefinition of $\Phi$ (leaving $\Psi$ unchanged): 
 \be
  \Phi\rightarrow \Phi' = \Phi-\frac{1}{2}\Psi+\frac{3}{2}\Delta^{-1}\ddot{\Psi}\,. 
 \ee
This implies 
 \be
  (4\Phi-2\Psi)\Delta \Psi=4\Phi'\Delta\Psi-6\Delta^{-1}\ddot{\Psi}\Delta \Psi\,, 
 \ee
 so that after cancelling the $\Delta$'s this  cancels the $\Psi\ddot{\Psi}$ term. 
 We obtain for the action in the new variables: 
 \be
 \begin{split}
  {\cal L}_{\rm FP} = &\,\frac{1}{4}\widehat{h}^{ij}\,\Box \,\widehat{h}_{ij} -\frac{1}{2}\Sigma^i \Delta \Sigma_i 
  +4\Phi'\Delta \Psi\\
  &+\frac{1}{2}\widehat{h}_{ij}\Pi^{ij} -\Sigma_i q^i-\Phi'\rho 
  - \frac{1}{2}\Psi\rho+\frac{3}{2}\dot{\Psi} q-3\Psi\, p\,, 
 \end{split}
 \ee
where we integrated by parts and used the conservation law $\Delta^{-1}\dot{\rho}=-q$. 
Integrating by parts again and using the last of (\ref{energymomentum}) this can also be rewritten as 
  \be
 \begin{split}
  {\cal L}_{\rm FP} = &\,\frac{1}{4}\widehat{h}^{ij}\,\Box \,\widehat{h}_{ij} -\frac{1}{2}\Sigma^i \Delta \Sigma_i 
  +4\Phi \Delta \Psi\\
  &+\frac{1}{2}\widehat{h}_{ij}\Pi^{ij} -\Sigma_i q^i-\Phi \rho - \frac{1}{2}\Psi(\rho+3 p-2\Delta \Pi)\,, 
 \end{split}
 \ee
where we  dropped the prime on $\Phi$. We could further  diagonalize the terms in the first line by defining 
 \be
  \Phi_{\pm} := 2\Psi \pm \Phi\,, 
 \ee
for which 
   \be
 \begin{split}
  {\cal L}_{\rm FP} = &\,\frac{1}{4}\widehat{h}^{ij}\,\Box \,\widehat{h}_{ij} -\frac{1}{2}\Sigma^i \Delta \Sigma_i 
  +\frac{1}{2}\Phi_+\Delta \Phi_{+}-\frac{1}{2}\Phi_{-}\Delta \Phi_{-}\\
  &+\frac{1}{2}\widehat{h}_{ij}\Pi^{ij} -\Sigma_i q^i
  -\frac{1}{2}(\Phi_{+}-\Phi_{-}) \rho - \frac{1}{8}(\Phi_{+}+\Phi_{-}) (\rho+3 p-2\Delta \Pi)\,. 
 \end{split}
 \ee
This is linearized gravity expressed in purely physical (gauge invariant) quantities.  
Let us finally note that we may integrate out $\Sigma_i$ and $\Phi_{\pm}$ by solving their own equations of motion, 
using the invertibility of $\Delta$.

\subsection{Homotopy interpretation of linearized gravity}\label{sectionhomotopyinterpretationlingrav}
We now interpret the gauge invariant variables for gravity in terms of homotopy transfer in the same way as for Maxwell's theory, 
focusing for simplicity on the vacuum case. 
We start from the chain complex defining linearized gravity 
  \be\label{chaimcomplexFP}
\begin{array}{ccccccccccc} X_1 &\xlongrightarrow{\partial} &X_0 {}  &\xlongrightarrow{\partial}
&X_{-1} &
\xlongrightarrow{\partial}& X_{-2}
\\[1.5ex]
\{\xi_{\mu}\}& &\{h_{\mu\nu}\} & &\{E_{\mu\nu}\}&
&\{F_{\mu}\}
\end{array}
\ee
where the spaces $X_0$ and $X_{-1}$ consist of symmetric but otherwise unconstrained Lorentz tensors (so that, for instance, $E_{\mu\nu}\in X_{-1}$ in general 
has a non-vanishing divergence). 
The differentials are defined by 
 \be\label{Abstrdifferential}
 \begin{split}
  \partial(\xi)_{\mu\nu}&=\partial_{\mu}\xi_{\nu}+\partial_{\nu}\xi_{\mu}\;, \\
  \partial(h)_{\mu\nu} &= G_{\mu\nu}(h)\,,\\
  \partial({E})_{\mu} &= \partial^{\nu}{E}_{\nu\mu}\,, 
 \end{split}
 \ee 
where $G_{\mu\nu}(h)$ is the linearized Einstein tensor  (\ref{linEinstein}). 
As before, the relation $\partial^2=0$ amounts to gauge invariance of $G_{\mu\nu}$ (when evaluated on $\xi\in X_1$) 
and to the Bianchi identity $\partial^{\mu}G_{\mu\nu}=0$ (when evaluated on $h\in X_0$).

In the same way as indicated in the diagram (\ref{DIAGRAM}) 
we want to map with  $p$ down to the space of gauge invariant fields $\bar{X}_0$, and $\bar{X}_1=\{0\}$ since 
there is then no gauge redundancy left. We write 
 \be
 p(h_{\mu\nu})=\bar{h}_{\mu\nu}=\begin{pmatrix} \bar{h}_{ij} & \bar{h}_{0i} \\ \bar{h}_{i0} &\bar{h}_{00} \end{pmatrix}\,,
 \ee  
 where in terms of the gauge invariant  variables 
  \be\label{intermsofBardeen}
   \bar{h}_{ij}=\widehat{h}_{ij}-2\Psi\delta_{ij}\;, \qquad \bar{h}_{0i}=-\Sigma_i\;, \qquad \bar{h}_{00}=-2\Phi \;. 
  \ee
Note that from these definitions it follows that 
  \be\label{gaugecond12}
  \partial^i\bar{h}_{ij}-\tfrac{1}{3}\partial_j(\bar{h}^{i}{}_{i})=0\,,\qquad \partial^i\bar{h}_{i0}=0\;. 
 \ee  
This is formally identical to the transverse gauge condition (but conceptually different because no gauge has been chosen). 
In order to identify the homotopy map so that the homotopy relation (\ref{homotopyRelation}) holds 
it is convenient to write the original field $h_{\mu\nu}$ in terms of the projected (gauge invariant) field plus a pure gauge term, 
 \be\label{hintobarh}
  h_{\mu\nu}=\bar{h}_{\mu\nu}+\partial_{\mu}A_{\nu}+\partial_{\nu}A_{\mu}\;, 
 \ee 
where 
 \be\label{Aintermsofh}
  A_{\mu}=(A_0,A_i)=(B-\dot{E}, E_i+\partial_iE)\;, 
 \ee 
as follows quickly with the formulas of the previous section.  
We now want to define a homotopy map $s: X_0\rightarrow X_1$  so that (\ref{homotopyRelation}) holds, for which 
we compute 
\be
   (\iota p-{\rm id})(h_{\mu\nu}) = \bar{h}_{\mu\nu}-h_{\mu\nu}=-\partial_{\mu}A_{\nu}-\partial_{\nu}A_{\mu} = -\partial(A)_{\mu\nu}\;, 
 \ee
where we recalled  that the inclusion map $\iota: \bar{X}_0\rightarrow X_0$ is simply  $\iota(\bar{h}_{\mu\nu})=\bar{h}_{\mu\nu}$. 
Assuming that the second term on the right-hand side of the homotopy relation (\ref{homotopyRelation}) vanishes due to $s_{-1}=0$
we infer 
 \be
  s_0(h)_{\mu}= A_{\mu}\in X_1\,, 
 \ee
where we mean that $A_{\mu}$ is defined in terms of the SVT-decomposed components of $h_{\mu\nu}$ as in (\ref{Aintermsofh}). 
Next, we have to check the homotopy relation on $\xi\in X_1$, which reads 
\be
  (\iota p-{\rm id})(\xi)=-\xi =-s_0(\partial \xi)\,,
\ee
since $p(\xi)=0$. This relation is satisfied since $\partial(\xi)$ by (\ref{Abstrdifferential}) is 
already of the pure gauge form and hence $s(\partial\xi)=\xi$.

Let us now determine the homotopy map on $X_{-2}$. We first need to specify how the projection map is extended to $X_{-1}$ and $X_{-2}$. 
For $E\in X_{-1}$ we set 
 \be
  \bar{E}_{\mu\nu}\equiv p({E})_{\mu\nu}=
  \begin{pmatrix} {E}_{00} & {E}_{0j}-\partial_j\Delta^{-1}(\partial^{\nu}{E}_{\nu0}) \\  {E}_{i0}-\partial_i\Delta^{-1}(\partial^{\nu}{E}_{\nu0})
  &{E}_{ij} -2\partial_{(i} \Delta^{-1}(\partial^{\nu}{E}_{j)\nu})+\partial_{i}\partial_{j}\Delta^{-2}(\partial^{\mu}\partial^{\nu}{E}_{\mu\nu})\end{pmatrix} . 
 \ee
This implements the projection to divergence-free Lorentz tensors:
 \be
  \partial^{\mu}\bar{E}_{\mu\nu}\equiv 0\;, 
 \ee 
and so $\bar{X}_{-1}$ consistent of symmetric tensors with vanishing divergence. Consequently, there is no non-trivial differential to $\bar{X}_{-2}$ 
and hence it is natural to assume that this space trivializes. 
Thus, we set for $F\in X_{-2}$  
 \be\label{p(F)=0}
  p(F)=0\;. 
 \ee
We now claim that the  homotopy map from $X_{-2}$ to $X_{-1}$ acting on $F_{\mu}=(F_0,F_i)$ is given by
 \be\label{homotopyFFF}
  s(F)_{\mu\nu}=\begin{pmatrix} s(F)_{00} & s(F)_{0j}\\ s(F)_{i0} & s(F)_{ij}\end{pmatrix}
  =\begin{pmatrix} 0  & \partial_j\Delta^{-1}F_0\\ \partial_i\Delta^{-1}F_0 & 2\partial_{(i}\Delta^{-1}F_{j)}-\partial_i\partial_j\Delta^{-2}(\partial_{\mu}F^{\mu})\end{pmatrix}.
 \ee
Let us verify the homotopy relation on ${E}\in X_{-1}$. We compute 
 \be
 \begin{split}
  (\iota p-{\rm id})({E})_{\mu\nu}&= \begin{pmatrix} 0 & -\partial_i\Delta^{-1}(\partial^{\nu}{E}_{\nu0}) \\  -\partial_i\Delta^{-1}(\partial^{\nu}{E}_{\nu0})
  & -2\partial_{(i} \Delta^{-1}(\partial^{\nu}{E}_{j)\nu})+\partial_{i}\partial_{j}\Delta^{-2}(\partial^{\mu}\partial^{\nu}{E}_{\mu\nu})\end{pmatrix} \\
  &= -s(\partial{E})_{\mu\nu}\;, 
 \end{split}
 \ee
where the second equality follows with (\ref{homotopyFFF}). 
Since  ${s}({E})=0$ this establishes the homotopy relation. 
Finally, for the homotopy relation on $F\in X_{-2}$ we have  with (\ref{p(F)=0}) 
 \be
   (\iota p-{\rm id})(F)_{\mu}=-F_{\mu}\,,
 \ee 
while we compute for the right-hand side, using $S(\partial(F))=0$, 
 \be
 \begin{split}
  -\partial(s(F))_{\mu}=-\partial^{\nu}s(F)_{\nu\mu}&=-\begin{pmatrix} \partial^0 s(F)_{00} + \partial^{i}s(F)_{i0}\\ \partial^0s(F)_{0i}+\partial^js(F)_{ji}\end{pmatrix} \\
  &=\begin{pmatrix} -F_0 \\ -\partial_i\Delta^{-1}(\partial^0F_0)-\partial_i\Delta^{-1}(\partial^jF_j)-F_i+\partial_i\Delta^{-1}(\partial_{\mu}F^{\mu})\end{pmatrix} \\
  &=\begin{pmatrix} -F_0 \\ -F_i\end{pmatrix}\,. 
 \end{split} 
 \ee 
This establishes  the homotopy relation and completes our discussion of the homotopy interpretation of gauge invariant variables 
in linearized gravity.

\section{Gauge invariant closed string theory} 

In this section we analyze the free action for the massless fields of closed bosonic string theory on a constant background in terms of gauge invariant field variables and 
show again that it can be interpreted as homotopy transfer. We employ the formalism of double field theory, subjected to a space/time split. 
This  theory  shows new features due to the chain complex being extended by a space of trivial parameters 
and a space of Bianchi identities for Bianchi identities (which in turn become important for the non-linear gauge algebra which is a genuine $L_{\infty}$ algebra).

\subsection{Linearized double field theory on flat backgrounds}

Our starting point is double field theory, which formulates the universal sector of the target space action of closed string theory, 
consisting of metric, B-field and dilaton, in terms of fields living  on a doubled space  \cite{Siegel:1993th,Hull:2009mi,Hohm:2010jy,Hohm:2010pp}. 
The doubling of coordinates is a feature of string theory on toroidal backgrounds, where they encode momentum and winding modes \cite{Kugo:1992md}, 
but as far as the free quadratic theory is concerned the doubling of coordinates is consistent for any background with constant metric and B-field. 
Since the free theory is all we need for the following discussion we will not elaborate on the physical meaning of the doubled coordinates any further. 

We use double field theory in a space/time split, as worked out by Naseer in \cite{Naseer:2015fba}, and expand the action  to quadratic order in fluctuations.  
The details of this computation will be detailed in a separate publication for the more general  case of time-dependent backgrounds, 
so here we only display the result. The field content is given by 
 \be
  h_{a\bar{b}}\;,\quad {\cal A}_{a}\;,\quad {\cal A}_{\bar{a}}\;, \quad \phi\;,\quad \varphi\;, 
 \ee
where $h_{a\bar{b}}$ encodes the spatial components of the metric and $B$-field fluctuations, the vectors ${\cal A}_{a}$ and ${\cal A}_{\bar{a}}$ encode 
(linear combinations of) the space-time component of the metric and $B$-field fluctuations, respectively, while $\varphi$ encodes the fluctuations of the dilaton 
and $\phi$ encodes the fluctuation of the time-time component of the metric (the lapse function). 
All fields depend on time $t$ and doubled spatial coordinates $x^M=(\tilde{x}_i, x^i)$, where $i=1,...,d$ and $M=1,...,2d$, with dual derivatives $\partial_M$. 
The dependence of the fields on doubled coordinates is subject to the level-matching constraint 
\begin{equation}
\label{weakconstraint}
\eta^{MN}\partial_M \partial_N f = 2 \tilde{\partial}^i \partial_i f  = 0 \,, \qquad \eta_{MN}  \ =  \ \begin{pmatrix} 0 & \delta^i{}_j  \\ \delta_i{}^j & 0 \end{pmatrix} \;, 
\end{equation}
where $f$ stands for any of the fields, and $\eta_{MN}$ is the $O(d,d)$ invariant metric that is used to raise and lower indices $M,N,\ldots$.  
The action is written in terms of the above fields and the background metric and $B$-field that are conveniently encoded in a constant background frame $E_{A}{}^{M}$, 
with flat index $A=(a,\bar{a})$, 
which is subject to the constraint that the following metric is block-diagonal: 
 \be\label{flatetaisG}
  {\cal G}_{AB} \ \equiv \ E_A{}^{M} E_{B}{}^{N} \eta_{MN} \ = \ \begin{pmatrix} -\delta_{ab} & 0 \\ 0 & \delta_{\bar{a}\bar{b}} \end{pmatrix}\;. 
 \ee
This metric is used to raise and lower flat indices $a,b,\ldots$ and $\bar{a},\bar{b},\ldots$. Furthermore, we define the derivative operators 
 \be
  D_A\equiv E_{A}{}^{M}\partial_M\equiv (D_a,D_{\bar{a}})\,,
 \ee
in terms of which the constraint (\ref{weakconstraint}) reads 
 \be\label{weakconstraintflat}
  \Delta\equiv -{2}D^a D_{a} =2 D^{\bar{a}} D_{\bar{a}}\;. 
 \ee
Thus, despite the doubling of spatial coordinates there is a unique Laplacian. 

We are now ready the write the quadratic double field theory action:  
 \be\label{quadrDFTaction}
  \begin{split}
   {\cal L}_{\rm DFT}=&-4(D_t\varphi)^2-D_{t}h^{a\bar{b}} D_{t}h_{a\bar{b}}\\
   &+8D^a\phi D_a\varphi -8 D^a\varphi D_a\varphi -8D_a\varphi D_{\bar{b}} h^{a\bar{b}}+4 D_a\phi D_{\bar{b}}h^{a\bar{b}}\\
   &-2D^{a} h^{b\bar{c}} D_{a}h_{b\bar{c}} +2 D^{c} h^{a\bar{b}} D_{a} h_{c\bar{b}} -2D^{\bar{c}}h^{a\bar{b}} D_{\bar{b}}h_{a\bar{c}}\;, 
  \end{split}
 \ee
where we defined the covariant time derivative operators 
 \be\label{covtimeder}
 \begin{split}
  D_t\varphi &= \partial_t\varphi +\frac{1}{2}D_a{\cal A}^{a}+\frac{1}{2}D_{\bar{a}}{\cal A}^{\bar{a}}\,,\\
  D_{t}h_{a\bar{b}} &= \partial_t h_{a\bar{b}}-D_a{\cal A}_{\bar{b}}+D_{\bar{b}}{\cal A}_{a}\,.
 \end{split} 
 \ee
 As unbarred and barred indices are consistently contracted the above action has a manifest $SO(d)_L\times SO(d)_R$ invariance, 
 a double copy structure inherited from the full string theory. 
 There is also an $O(d,d)$ \textit{duality}, i.e., an invariance under a simultaneous action on both the fields and the background. 
 We note that the action is the complete quadratic action for the massless string fields as there is a field basis for which the 
 higher-derivative $\alpha'$ corrections  become only relevant at cubic level. 
 The action is invariant under the gauge symmetries with parameters $(\xi_a,\xi_{\bar{a}},\xi^0,\xi_0)$, given by 
 \be\label{DFGgaugetrans}
  \begin{split}
   \delta h_{a\bar{b}} &= D_a\xi_{\bar{b}} -D_{\bar{b}}\xi_a\,,\\
   \delta \phi &=\dot{\xi}^0\,,\\
   \delta \varphi &= -\frac{1}{2}D_a\xi^a -\frac{1}{2}D_{\bar{a}} \xi^{\bar{a}}\,, \\
   \delta {\cal A}_{a} &=\dot{\xi}_a +D_{a}({\xi}_0+\xi^0)\,, \\
   \delta {\cal A}_{\bar{a}} &=\dot{\xi}_{\bar{a}} +D_{\bar{a}}({\xi}_0-\xi^0)\,. 
  \end{split}
 \ee 
There is a gauge symmetry for gauge symmetries since gauge parameters of the form 
 \be\label{trivialparameterchi}
  \xi_{a}=D_{a}\chi\;, \quad \xi_{\bar{a}}=D_{\bar{a}}\chi\;, \quad {\xi}_{0}=-\dot{\chi} \;, 
 \ee  
where $\chi$ is an arbitrary scalar, do not generate a  transformation of  fields. 
 Note that the derivatives (\ref{covtimeder}) are invariant under gauge transformations w.r.t.~$(\xi_a,\xi_{\bar{a}})$. 
They are also invariant under $\xi_0$ transformations thanks to the constraint (\ref{weakconstraintflat}). This leaves  
$\xi^0$ transformations as the only symmetry linking the terms in the first line of (\ref{quadrDFTaction}) to the rest of 
the action. The total invariance, subject to the level-matching constraint (\ref{weakconstraintflat}),  is easy to verify. 
It should be emphasized that at the level of the free theory one does not have to worry about how to implement the constraint (\ref{weakconstraintflat}) 
on products of fields, since under an integral it is sufficient that all fields and gauge parameters satisfy the constraint \cite{Hull:2009mi}.

Our goal in the following is to rewrite linearized double field theory in terms of gauge invariant variables. 
We start by subjecting the fields to a scalar-vector-tensor (SVT) decomposition as for conventional gravity, writing 
 \be
  \begin{split}
   h_{a\bar{b}}&=\widehat{h}_{a\bar{b}} +D_{a} B_{\bar{b}} -D_{\bar{b}}B_{a} + D_aD_{\bar{b}}E\,,\\
   {\cal A}_{a} &= A_{a}+D_{a}A\,,\\
   {\cal A}_{\bar{a}} &= A_{\bar{a}} +D_{\bar{a}}\bar{A}\,,
  \end{split}
 \ee
where 
 \be
  D^{a}\,\widehat{h}_{a\bar{b}}=D^{\bar{b}}\,\widehat{h}_{a\bar{b}}=0\;, \qquad D^{a}B_{a}=D^{\bar{a}}B_{\bar{a}} = 0 
  \;, \qquad D^{a}A_{a}=D^{\bar{a}}A_{\bar{a}} = 0 \,. 
 \ee
Note that, in contrast to conventional gravity, we have not taken out the trace part  of $\widehat{h}_{a\bar{b}}$ since 
there is no $SO(d)_L\times SO(d)_R$ invariant way to do so. 
Similarly, we write for the gauge parameter: 
 \be\label{SVTparameters}
  \xi_a=\zeta_a+D_{a}\lambda\,, \qquad 
  \xi_{\bar{a}} = \zeta_{\bar{a}}+D_{\bar{a}}\bar{\lambda}\,,
 \ee
where 
 \be
  D^{a}\zeta_{a}=D^{\bar{a}}\zeta_{\bar{a}} = 0\,. 
 \ee
It is straightforward to work out the gauge transformations of the SVT components: 
 \be
  \begin{split}
   \delta \widehat{h}_{a\bar{b}} &= 0\,,\qquad 
   \delta B_{a} = \zeta_a\,,\qquad 
   \delta B_{\bar{a}} = \zeta_{\bar{a}}\,,\qquad \qquad \qquad 
   \delta E=\bar{\lambda}-\lambda\,,\\
   \delta A_a & = \dot{\zeta}_a\,,\quad \;\;
   \delta A_{\bar{a}} = \dot{\zeta}_{\bar{a}}\,,\qquad \;\;
   \delta A = \dot{\lambda}+\tilde{\xi}_0 +\xi^0\,,\qquad 
    \delta \bar{A} = \dot{\bar{\lambda}}+\tilde{\xi}_0 -\xi^0\,,\\
    \delta \varphi &= \frac{1}{4}\Delta (\lambda - \bar{\lambda})\,,\qquad \qquad \quad 
     \delta \phi =\dot{\xi}^0\,. 
  \end{split}
 \ee  
We observe that $\widehat{h}_{a\bar{b}}$ is gauge invariant. In addition, it is easy to see that one may define 
two gauge invariant vector modes and two gauge invariant scalar modes: 
 \be\label{gaugeinvDFTfields}
  \begin{split}
   \widehat{A}_{a} &= A_{a}-\dot{B}_{a}\,,\\
    \widehat{A}_{\bar{a}} &= A_{\bar{a}}-\dot{B}_{\bar{a}}\,, \\
   \Psi &= \phi-\frac{1}{2}\frac{d}{dt}(A-\bar{A}) - \frac{1}{2}\ddot{E}\,, \\
   \Phi &= \varphi + \frac{1}{4}\Delta E\,. 
  \end{split}
 \ee 
Thus, compared to pure gravity (c.f.~(\ref{gaueinvgravitymodes})) we have one more vector mode, 
which corresponds to the vector component of the B-field.  
Somewhat surprisingly, however, we do not obtain an additional scalar mode corresponding to the dilaton. 
In order to clarify this point let us count the total number of off-shell field components and of gauge invariant quantities, 
taking  for definiteness $d=3$.  There are then 17 off-shell field components distributed as 
\be
  h_{a\bar{b}}\,: \;\; 9\, \quad {\cal A}_{a}\,: \;\; 3\,  \quad {\cal A}_{\bar{a}}\,: \;\; 3\, \quad \phi\,: \;\; 1\, \quad\varphi\,: \;\; 1\,. 
 \ee 
Naively there are eight gauge redundancies ($3+3$ for $\xi_a$ and $\xi_{\bar{a}}$ and $1+1$ for ${\xi}_0$ and $\xi^0$), 
but there is the one-parameter gauge symmetry of gauge symmetries (\ref{trivialparameterchi}), leaving seven gauge redundancies of fields. 
We thus expect ten gauge invariant combinations, and indeed the counting works with the above gauge invariant variables: 
 \be
  10  \ =  \ 4 ( \widehat{h}_{a\bar{b}}) \ + \ 2 (\widehat{A}_{a})\ + \ 2 (\widehat{A}_{\bar{a}}) \ + \ 1 (\Psi) \ + \ 1 (\Phi)\,. 
 \ee
Note, in particular, that $\widehat{h}_{a\bar{b}}$ encodes in a gauge invariant way all four propagating degrees of freedom, i.e., 
the spin-2 tensor modes with two degrees of freedom, the scalar mode given by the B-field and the scalar mode given by the dilaton.

We now return to the quadratic double field theory Lagrangian (\ref{quadrDFTaction}) and express it in terms of the gauge invariant variables. 
It is a straightforward computation to verify that the Lagrangian can be written as  
  \be
   {\cal L}=-4\dot{\Phi}^2 - \widehat{h}^{a\bar{b}}\,\Box\, \widehat{h}_{a\bar{b}} +\frac{1}{2}\widehat{A}^{a}\Delta \widehat{A}_{a}
   -\frac{1}{2}\widehat{A}^{\bar{a}}\Delta \widehat{A}_{\bar{a}}
   -4\Phi\Delta \Phi +4 \Psi\Delta \Phi\,, 
  \ee
where $\Box = -\frac{\partial^2}{\partial t^2}+\Delta$ is the d'Alembert operator. 
As encountered before for conventional gravity, 
the action is not diagonal and includes a term with time derivatives for $\Phi$. This does not mean that $\Phi$ is propagating, 
however, since we may perform the following  field redefinition 
 \be
  \Psi\rightarrow \Psi' = \Psi -\Phi+\Delta^{-1}\ddot{\Phi}\,, 
 \ee 
which removes the $\dot{\Phi}^2$ term, 
after which the Lagrangian reads 
 \be
   {\cal L}= - \widehat{h}^{a\bar{b}}\,\Box\, \widehat{h}_{a\bar{b}} +\frac{1}{2}\widehat{A}^{a}\Delta \widehat{A}_{a}-\frac{1}{2}\widehat{A}^{\bar{a}}\Delta \widehat{A}_{\bar{a}}
    +4\Psi'\Delta \Phi\;. 
 \ee 
This form makes it manifest that only $\widehat{h}^{a\bar{b}}$ propagates, while the other modes could be integrated out to eliminate them, 
i.e., in the vacuum case the Lagrangian reduces to ${\cal L}= - \widehat{h}^{a\bar{b}}\,\Box\, \widehat{h}_{a\bar{b}}$.

\subsection{Chain complex of double field theory}

In this subsection we set the stage for the interpretation of gauge invariant variables in double field theory  in terms of homotopy transfer 
by defining the chain complex encoding the free  theory. 
This brings in some conceptually new features due to the chain complex being extended compared to Yang-Mills and gravity theories. 
The chain complex  reads 
  \be\label{chaimcomplexDFT}
\begin{array}{ccccccccccc} X_2 &\xlongrightarrow{\partial} & X_1 &\xlongrightarrow{\partial} &X_0 {}  &\xlongrightarrow{\partial}
&X_{-1} &
\xlongrightarrow{\partial}& X_{-2} & \xlongrightarrow{\partial}& X_{-3}
\\[1.5ex]
\{\chi\}& & \{\xi\}& &\{\Psi\} & &\{{\cal E}\} & & \{{\cal G}\}  & & \{\rho \} 
\end{array}
\ee
We have to define each vector space and each differential map: 
$X_2$ is the space of trivial gauge parameters, which here are scalars $\chi$. $X_{1}$ is the space of gauge parameters $\xi$, 
and $X_0$ is the space of fields $\Psi$, which read in terms of field components  
 \be\label{manycomponentobjects}
   \xi= \begin{pmatrix} \xi^{a} \\ \xi^{\bar{a}} \\ \xi_0 \\ \xi^0 \end{pmatrix}\in X_1\;, \qquad 
    \Psi=\begin{pmatrix} h_{a\bar{b}} \\ \phi \\ \varphi \\ {\cal A}_{a} \\ {\cal A}_{\bar{a}} \end{pmatrix}\in X_0 \;. 
 \ee
Next, $X_{-1}$ is the space of field equations ${\cal E}$, which have the same index structure as fields, and $X_{-2}$ is the space of Bianchi identities, 
which have the same index structure as the gauge parameters. Finally, $X_{-3}$ is the space of Bianchi identities for Bianchi identities, which 
have the same index structure as the trivial gauge parameters
and are thus scalars. More abstractly, we have the following equivalence  between vector spaces and their duals:   
 \be
  X_{-1}\cong (X_0)^{\star}\,,\quad  X_{-2}\cong (X_1)^{\star} \,, \quad X_{-3}\cong (X_2)^{\star}\,. 
 \ee

Let us now give the differential maps $\partial$. From the form of the trivial gauge parameters (\ref{trivialparameterchi}) we read off 
for  $\partial_2 : X_2\rightarrow X_1$, 
 \be\label{triviailfortrivialtrans}
  \partial(\chi)=\begin{pmatrix} D_a\chi \\ D_{\bar{a}}\chi \\ -\dot{\chi} \\ 0\end{pmatrix} \in X_1\;. 
 \ee
Next, the gauge transformations (\ref{DFGgaugetrans}) determine the differential $\partial_1 :X_1\rightarrow X_0$ via 
$\delta_{\xi}\Psi=\partial(\xi)$: 
 \be
  \begin{split}
  \partial(\xi)=\begin{pmatrix} D_a\xi_{\bar{b}} -D_{\bar{b}}\xi_a \\ \dot{\xi}^0\\ -\frac{1}{2}D_a\xi^a -\frac{1}{2}D_{\bar{a}} \xi^{\bar{a}} 
  \\ \dot{\xi}_a +D_{a}({\xi}_0+\xi^0) \\ \dot{\xi}_{\bar{a}} +D_{\bar{a}}({\xi}_0-\xi^0) \end{pmatrix}\in X_0\,. 
  \end{split}
 \ee 
The field equations ${\cal E}=0$ are encoded in the differential  $\partial_0 :X_0\rightarrow X_{-1}$ via 
 \be
     \begin{pmatrix} {\cal E}_{a\bar{b}} \\ {\cal E}_\phi \\ {\cal E}_\varphi \\ {\cal E}_{a} \\ {\cal E}_{\bar{a}} \end{pmatrix} 
     \ = \ \partial \begin{pmatrix} h_{a\bar{b}} \\ \phi \\ \varphi \\ {\cal A}_{a} \\ {\cal A}_{\bar{a}} \end{pmatrix} \in X_{-1}
   \,, 
 \ee
for which one obtains upon variation of (\ref{quadrDFTaction}) 
 \be\label{EQuationsOFMotion}
  \begin{split}
   {\cal E}_{a\bar{b}}&=\partial_t(D_th_{a\bar{b}})-2D_aD_{\bar{b}}\phi+{\cal R}_{a\bar{b}}\,,\\
   {\cal E}_{\phi} &= {\cal R}\;,\\
   {\cal E}_{\varphi} &= \partial_t(D_t\varphi)+\frac{1}{2}\Delta\phi+{\cal R}\,,\\
   {\cal E}_{a} &= D_{a}(D_t\varphi)  +\frac{1}{2}D^{\bar{b}}(D_{t}h_{a\bar{b}})\,,\\
   {\cal E}_{\bar{a}} &= D_{\bar{a}}(D_t\varphi)-\frac{1}{2}D^{b}(D_t h_{b\bar{a}})\,, 
  \end{split}
 \ee  
 with the spatial  curvatures 
  \be\label{SpatialCurv}
   \begin{split}
    {\cal R} &= -\Delta\varphi + D_{a}D_{\bar{b}} h^{a\bar{b}}\,,\\
    {\cal R}_{a\bar{b}} &= 2\left(-\frac{1}{2}\Delta h_{a\bar{b}}-D_{a}D^{c} h_{c\bar{b}}+D_{\bar{b}}D^{\bar{c}} h_{a\bar{c}}+2 D_{a}D_{\bar{b}}\varphi\right)\;. 
   \end{split}
  \ee 
The gauge invariance of (\ref{EQuationsOFMotion}), and thus the nilpotency $\partial_0\circ \partial_1=0$, may be easily verified, using in particular
 \be
  \delta(D_t h_{a\bar{b}})=2 D_{a}D_{\bar{b}}\xi^{0}\;, \qquad \delta(D_t\varphi)=-\frac{1}{2}\Delta \xi^0\;. 
 \ee  
We next observe that the curvatures (\ref{SpatialCurv}) obey the Bianchi identities  
  \be
  \begin{split}
   &D^{a}{\cal R}_{a\bar{b}}-2D_{\bar{b}}{\cal R}=0\,,\\
   &D^{\bar{b}}{\cal R}_{a\bar{b}}+2D_{a}{\cal R}=0\,, 
  \end{split}
 \ee 
from which in turn one may derive Bianchi identities for the ${\cal E}$. 
These identities are encoded in the next differential $\partial_{-1}: X_{-1}\rightarrow X_{-2}$: 
 \be\label{GgaugeBianchi}
 \partial{\cal E} 
 \equiv 
   \begin{pmatrix} (\partial{\cal E})^{a} \\ (\partial{\cal E})^{\bar{a}} \\ (\partial{\cal E})_0 \\ (\partial{\cal E})^0 \end{pmatrix} \equiv
  \partial 
  \begin{pmatrix} {\cal E}_{a\bar{b}} \\ {\cal E}_\phi \\ {\cal E}_\varphi \\ {\cal E}_{a} \\ {\cal E}_{\bar{a}} \end{pmatrix}
  = \begin{pmatrix}D_{\bar{b}}{\cal E}^{a\bar{b}}+2D^a{\cal E}_{\varphi}-2\partial_t{\cal E}^{a} \\ 
  D_{b}{\cal E}^{b\bar{a}}-2D^{\bar{a}}{\cal E}_{\varphi}+2\partial_t{\cal E}^{\bar{a}} \\ 
  D_{a}{\cal E}^{a}+D_{\bar{a}}{\cal E}^{\bar{a}} \\ 
  D_{a}{\cal E}^{a}-D_{\bar{a}}{\cal E}^{\bar{a}}-\partial_t{\cal E}_{\phi}\,
   \end{pmatrix} \,.
 \ee 
Indeed, for ${\cal E}$ given by (\ref{EQuationsOFMotion}) these combinations vanish identically, which proves  $\partial_{-1}\circ\partial_{0}=0$.  
Finally, the differential $\partial_{-2}: X_{-1}\rightarrow X_{-3}$, defined by 
 \be\label{partialonCalG}
  \partial  \begin{pmatrix} {\cal G}^{a} \\ {\cal G}^{\bar{a}} \\ {\cal G}_0 \\ {\cal G}^0 \end{pmatrix}\equiv 
  \dot{\cal G}_0+\frac{1}{2}D_{a}{\cal G}^{a}-\frac{1}{2}D_{\bar{a}}{\cal G}^{\bar{a}}\;, 
 \ee
 encodes Bianchi identities of Bianchi identities. Indeed, for ${\cal G}$'s given by  the right-hand side of (\ref{GgaugeBianchi}) this vanishes identically, 
irrespective of what the ${\cal E}$'s are, establishing $\partial_{-2}\circ\partial_{{-1}}=0$. 
This completes our discussion of the chain complex encoding the gauge structure and dynamics of linearized double field theory on flat space.

\subsection{Homotopy interpretation}

We now turn to the homotopy transfer from the chain complex (\ref{chaimcomplexDFT}) to a projected 
complex in which in particular the fields are gauge invariant and so the gauge symmetries  trivialize. 
Since we now also have gauge for gauge symmetries it is in principle conceivable that the space of gauge parameters 
is projected to something non-zero but trivial, but we will see in the following that the natural  homotopy relations 
work with 
 \be
  p_{2}=p_{1}=0\;,
 \ee
and so we indeed take $\bar{X}_{2}=\{0\}=\bar{X}_{1}$. 
In order to define the homotopy map $s_1:X_1\rightarrow X_{2}$ we follow a similar strategy as for the 
fields in Maxwell's theory and standard gravity. We first decompose the gauge parameters into combinations 
that are invariant under the gauge for gauge transformations $\delta_{\chi}\xi=\partial(\chi)$. Using the SVT decomposition (\ref{SVTparameters}) 
for the gauge parameters and (\ref{triviailfortrivialtrans}) one finds 
 \be
 \begin{split}
  \delta_{\chi}\zeta^{a}&=0\;, \qquad \delta_{\chi}\zeta^{\bar{a}}=0\;, \\
  \delta_{\chi}\lambda&=\chi\;, \qquad \delta_{\chi}\bar{\lambda}=\chi\;, \\
  \delta_{\chi}\xi_0&=-\dot{\chi}\;, \quad \delta_{\chi}\xi^{0}=0\;. 
 \end{split} 
 \ee
We can then define the particular combination of gauge for gauge  invariant parameters 
 \be\label{gaugeforgaugeinv}
  \begin{split}
   \bar{\xi}^{a}&\equiv\zeta^{a}+\frac{1}{2}D^{a}(\lambda-\bar{\lambda})\;, \\
   \bar{\xi}^{\bar{a}}&\equiv\zeta^{\bar{a}}-\frac{1}{2}D^{\bar{a}}(\lambda-\bar{\lambda})\;,\\
   \bar{\xi}_0 &\equiv\xi_0 +\frac{1}{2}(\dot{\lambda}+\dot{\bar{\lambda}})\;, \\
   \bar{\xi}^0 &\equiv \xi^{0}\;. 
  \end{split}
 \ee 
With  these we can write a general decomposition of a gauge parameter into
a gauge for gauge invariant parameter plus the trivial or `pure gauge for gauge' part,   
 \be\label{homotopoyXidecomposition}
   \xi= \begin{pmatrix} \xi^{a} \\ \xi^{\bar{a}} \\ \xi_0 \\ \xi^0 \end{pmatrix}
   = \begin{pmatrix} \bar{\xi}^{a}+D^{a}f \\ \bar{\xi}^{\bar{a}} +D^{\bar{a}}f \\ \bar{\xi}_0 -\dot{f}\\ \bar{\xi}^0 \end{pmatrix}\;. 
 \ee 
A quick computation confirms that this holds identically for  
 \be
  f\equiv \frac{1}{2}(\lambda+\bar{\lambda})\;. 
 \ee  
It is convenient to write (\ref{homotopoyXidecomposition}) more briefly as 
 \be\label{decompo}
  \xi=\bar{\xi}+\partial(f)\;. 
 \ee 
In analogy to previous cases it is natural to define the homotopy map $s$ acting on gauge parameters as 
 \be\label{SofXII}
  s(\xi)=f\;. 
 \ee   
The homotopy relation on $X_2$ then follows quickly, using $s(\chi)=0$: 
 \be
  ({\rm id}-\iota p)(\chi) = \chi = s(\partial\chi)\;. 
 \ee

Next, we define the homotopy $s_0:X_0\rightarrow X_{1}$ from fields to gauge parameters, 
using again the strategy of decomposing  a field into a gauge invariant part plus a pure gauge part.  
Denoting the gauge invariant quantities with a bar for consistency of notation, we rename 
(\ref{gaugeinvDFTfields}) into 
 \be
  \bar{h}_{a\bar{b}}=\widehat{h}_{a\bar{b}}\;, \quad \bar{\varphi}=\Phi\;, \quad \bar{\phi}=\Psi\;, \quad \bar{{\cal A}}_{a}=\widehat{A}_{a}\;, \quad 
  \bar{{\cal A}}_{\bar a}=\widehat{A}_{\bar a}\;. 
 \ee 
These define the projection map $p_0: X_0\rightarrow \bar{X}_0$, 
 \be
  p_0(\Psi)=\bar{\Psi}\,,
 \ee 
using the notation (\ref{manycomponentobjects}) (and its obvious extension to $\bar{X}_0$). 
It is then a straightforward computation to verify that the original fields are related to the gauge invariant fields via 
 \be\label{generalDFTdecomp}
 \begin{split}
  h_{a\bar{b}}&=\bar{h}_{a\bar{b}}+D_{a}F_{\bar{b}}-D_{\bar{b}}F_{a}\;,\\
  \phi &= \bar{\phi}+\dot{F}^{0}\;, \\
  \varphi &= \bar{\varphi}-\frac{1}{2}D_{a}F^{a}-\frac{1}{2}D_{\bar{a}}F^{\bar{a}}\;, \\
  {\cal A}_{a}&=\bar{\cal A}_{a}+\dot{F}_{a}+D_{a}(\widetilde{F}_{0}+F^{0})\,,\\
  {\cal A}_{a}&=\bar{\cal A}_{\bar{a}} +\dot{F}_{\bar{a}}+D_{\bar{a}}(\widetilde{F}_{0}-F^0)\;, 
 \end{split}
 \ee 
where the `pure gauge'  $F$ terms are given in terms of the SVT components of the fundamental fields by 
 \be
  \begin{split}
   F_{a}&=B_{a}-\frac{1}{2}D_{a}E\;, \\
   F_{\bar{a}} &= B_{\bar{a}}+\frac{1}{2}D_{\bar{a}}E\;, \\
   F^{0} &= \frac{1}{2}(\dot{E}+A-\bar{A})\;, \\
   \widetilde{F}_{0}&=\frac{1}{2}(A+\bar{A})\;. 
  \end{split}
 \ee  
We can also write this relation compactly as $\Psi=\bar{\Psi}+\partial F$.  
These define the homotopy map $s$ from the space of fields to the space of gauge parameters, 
  \be\label{SofPSIII}
  s(\Psi) =  \begin{pmatrix} F^{a} \\ F^{\bar{a}} \\ \widetilde{F}_0 \\ F^0 \end{pmatrix}\;. 
 \ee
From these formulas it follows by a direct computation that $s$ acting on a field that is pure gauge gives back 
the gauge for gauge invariant part (\ref{gaugeforgaugeinv}) of the parameter:
 \be
  s(\partial \xi)=\bar{\xi}\;. 
 \ee
The homotopy relation evaluated on $\xi\in X_{1}$ then follows with (\ref{decompo}) 
 \be
  ({\rm id}-\iota p)(\xi)=\xi=\partial (s(\xi))+s(\partial\xi)=\partial f+\bar{\xi}\,, 
 \ee
where we used (\ref{SofXII}). 
Finally, we can inspect  the homotopy relation for $\Psi\in X_0$, 
 \be
  ({\rm id}-\iota p)(\Psi)=\Psi-\bar{\Psi} = \partial(F)=\partial(s(\Psi))\,,
 \ee
where we used (\ref{generalDFTdecomp}) and (\ref{SofPSIII}). 
This shows that the homotopy relation is satisfied if we take the homotopy to be trivial on 
the space of field equations, 
 \be
  s_{-1}=0\,. 
 \ee
Summarizing, on the sub-complex $X_2\rightarrow X_1\rightarrow X_0$ the maps (\ref{SofXII}) and (\ref{SofPSIII}) 
define a consistent homotopy transfer.

In the remainder of this section we show that the homotopy transfer can be extended to the entire chain complex, 
which requires two new non-trivial homotopy maps. As in the Yang-Mills and gravity examples, however, the spaces of Bianchi (and Bianchi for Bianchi) identities 
will be projected to zero: 
 \be
  \bar{X}_{-2}=\bar{X}_{-3}=\{0\}\;, 
 \ee
i..e, we set $p_{-2}=p_{-3}=0$. The projector $p_{-1}:X_{-1}\rightarrow \bar{X}_{-1}$ is non-trivial, however, and projects onto the space 
of tensors of the equations-of-motion type that satisfy the Bianchi identities identically. 
Writing for the projected tensors  $\bar{\cal E}=p({\cal E})$ one finds 
 \be\label{projecgorcalE}
  \begin{split}
   \bar{\cal E}_{a\bar{b}} &= {\cal E}_{a\bar{b}}+2 D_{a}\Delta^{-1}(\partial {\cal E})_{\bar{b}}-2D_{\bar{b}}\Delta^{-1}(\partial {\cal E})_{a}\\
   &\qquad \;\;\, +4 D_a D_{\bar{b}} \Delta^{-2}\left((\dot{\partial{\cal E}})^0-\frac{1}{2} D^c(\partial{\cal E})_c -\frac{1}{2} D^{\bar{c}}(\partial{\cal E})_{\bar{c}}\right)\,,\\
   \bar{\cal E}_{\phi} &= {\cal E}_{\phi}\,,\\
   \bar{\cal E}_{\varphi} &= {\cal E}_{\varphi}\,, \\
   \bar{\cal E}_{a} &= {\cal E}_{a}+ D_{a}\Delta^{-1}\left(({\partial{\cal E}})_0  + ({\partial{\cal E}})^0\right)\,,\\
   \bar{\cal E}_{\bar{a}} &= {\cal E}_{\bar{a}} -D_{\bar{a}}\Delta^{-1}\left(({\partial{\cal E}})_0  - ({\partial{\cal E}})^0\right)\,, 
  \end{split} 
 \ee
where the components of $(\partial{\cal E})$ are defined in (\ref{GgaugeBianchi}). 
This is the appropriate projector since one may now verify that  
the following relations hold identically, just as a consequence of the definition of the bar: 
\be
\begin{split}
0 &\equiv D_{\bar{b}}\bar{\cal E}^{a\bar{b}}+2D^a\bar{\cal E}_{\varphi}-2\partial_t\bar{\cal E}^{a} \,, \\
0 &\equiv  D_{b}\bar{\cal E}^{b\bar{a}}-2D^{\bar{a}}\bar{\cal E}_{\varphi}+2\partial_t\bar{\cal E}^{\bar{a}}\,, \\ 
0 &\equiv  D_{a}\bar{\cal E}^{a}+D_{\bar{a}}\bar{\cal E}^{\bar{a}}\,, \\ 
0 &\equiv  D_{a}\bar{\cal E}^{a}-D_{\bar{a}}\bar{\cal E}^{\bar{a}}-\partial_t\bar{\cal E}_{\phi}\,. 
\end{split}
\ee

We have now completed the definition of the projection of the entire chain complex. It remains to identify the 
homotopy maps. We continue to take the homotopy from the space of field equations to the space of fields to be trivial, $s({\cal E})=0$. 
In order to find the homotopy map on the space of Bianchi identities we use this to evaluate 
the homotopy relation on ${\cal E}\in X_{-1}$: 
 \be
  ({\rm id}-\iota p)({\cal E}) = {\cal E}-\bar{\cal E}=s(\partial {\cal E})\;. 
 \ee
Since we expressed the projector  (\ref{projecgorcalE}) already in terms of $\partial{\cal E}$ we can immediately read off 
the homotopy map so that the homotopy  relation holds: 
 \be\label{homotopyonG}
 \begin{split}
  s({\cal G})_{a\bar{b}} &= -2 D_{a}\Delta^{-1}{\cal G}_{\bar{b}}+2D_{\bar{b}}\Delta^{-1}{\cal G}_{a} 
   -4 D_a D_{\bar{b}} \Delta^{-2}\left(\dot{{\cal G}}^0-\frac{1}{2} D^c{\cal G}_c -\frac{1}{2} D^{\bar{c}}{\cal G}_{\bar{c}}\right)\,,\\
   s({\cal G})_{\phi} &=0\,,\\
   s({\cal G})_{\varphi} &= 0\,, \\
   s({\cal G})_a &= -D_a\Delta^{-1}\left({\cal G}_0+{\cal G}^0\right)\,,\\
   s({\cal G})_{\bar{a}} &= D_{\bar{a}}\Delta^{-1}\left({\cal G}_0- {\cal G}^0\right)\,. 
 \end{split}
 \ee
 
Let us next turn to the homotopy relation  on ${\cal G}\in  X_{-2}$. 
 Recalling $p({\cal G})=0$, it reads 
 \be\label{HomotopyOnCALLLG} 
   ({\rm id}-\iota p)({\cal G}) = {\cal G} = \partial(s({\cal G}))+s(\partial{\cal G})\;. 
 \ee 
The first term on the right-hand side can be computed directly with (\ref{GgaugeBianchi}) and (\ref{homotopyonG}), for which 
one finds 
 \be
  \partial(s({\cal G}))={\cal G}+2\begin{pmatrix} D_a\Delta^{-1}\big(\dot{\cal G}_0+\tfrac{1}{2}D^c{\cal G}_c-\tfrac{1}{2}D^{\bar{c}}{\cal G}_{\bar{c}}\big) \\ 
  D_{\bar{a}}\Delta^{-1}\big(\dot{\cal G}_0+\tfrac{1}{2}D^c{\cal G}_c-\tfrac{1}{2}D^{\bar{c}}{\cal G}_{\bar{c}}\big) \\
  0 \\ 0\end{pmatrix}\,. 
 \ee
Looking back at (\ref{partialonCalG}) we infer that the failure of  $\partial(s({\cal G}))$ to give back ${\cal G}$ involves $\partial {\cal G}$. 
This implies that (\ref{HomotopyOnCALLLG}) is satisfied if we define a non-trivial homotopy map $s:X_{-3}\rightarrow X_{-2}$ as follows 
 \be\label{SofRHO}
  s(\rho) = -2\begin{pmatrix} D_a\Delta^{-1}\rho \\ 
  D_{\bar{a}}\Delta^{-1}\rho \\
  0 \\ 0\end{pmatrix}\;. 
 \ee
Finally, it then remains to check that the homotopy relation holds on $\rho\in X_{-3}$: 
  \be
   ({\rm id}-\iota p)(\rho) = \rho = \partial(s(\rho))\;, 
  \ee
using $\partial\rho=0$.  This identity is quickly verified with (\ref{partialonCalG}) and (\ref{SofRHO}). 
With this we have completed the proof that the entire chain complex (\ref{chaimcomplexDFT}) of double field theory 
can be homotopy transferred to the complex $\bar{X}_{0}\rightarrow \bar{X}_{-1}$ of gauge invariant fields and their field 
equations, where all redundancies encoded in gauge symmetries and Bianchi identities have been eliminated.

\section{Homotopy Transfer} 

In this section we introduce the general framework of $L_{\infty}$-algebras and homotopy transfer and explain how it 
determines gauge invariant variables. 
We begin by giving a brief but hopefully self-contained introduction to the coalgebra formulation that is most useful for 
our purposes. In the second subsection we give the details of the perturbation lemma, and in the third subsection 
we illustrate the resulting techniques for the example of Yang-Mills theory. 

\subsection{General theory}

\subsubsection*{The symmetric coalgebra}

A convenient formulation of $L_\infty$-algebras makes use of the notions of coalgebras and coderivations. This construction can be found for example in \cite{Lada:1994mn}. The very similar construction for $A_\infty$-algebras is nicely described in \cite{Kajiura2003}. We only need one type of coalgebra. Given a graded vector space $X_\bullet$ (the total vector space of the chain complex), 
we can define the symmetric graded coalgebra $S^c(X_\bullet)$. As a vector space, we can think of it as the Fock space made from elements in $X_\bullet$, where elements of odd degree are fermionic (anti-commuting), and elements of even degree are bosonic (commuting). For our purposes, we can exclude the zero particle state. We therefore define
\begin{equation}
S^c(X_\bullet) = \bigoplus_{n \ge 1} X_\bullet^{\wedge n}.
\end{equation}
The wedge denotes the graded symmetric product of vector spaces.

The structure turning $S^c(X_\bullet)$ into a coalgebra is the coproduct
\begin{equation}
\Delta: S^c(X_\bullet) \rightarrow S^c(X_\bullet) \otimes S^c(X_\bullet).
\end{equation}
Notice that the direction of the map $\Delta$ is opposite from what we want from an ordinary product. It takes one input and gives two outputs. The coproduct is the sum of all possible ways to split an element in $S^c(X_\bullet)$ into two in a symmetric fashion. We give a few examples before defining it in general. When $\Delta$ acts on a single element 
$x \in X_\bullet \subseteq S^c(X_\bullet)$ there is no sensible way to split it into two pieces, and  we therefore define
\begin{equation}
\Delta(x) = 0.
\end{equation}
Next we can consider $x_1 \wedge x_2 \in X_\bullet \wedge X_\bullet \subseteq S^c(X_\bullet)$. In that case, the obvious split reads
\begin{equation}
\Delta(x_1 \wedge x_2) = x_1 \otimes x_2 + (-)^{x_1 x_2} x_2 \otimes x_1.
\end{equation} 
The sum makes sure that $\Delta$ is well defined on $S^c(X_\bullet)$, i.e.~that it is invariant under interchanging $x_1$ and $x_2$, up to a sign. As a final example, we consider a cubic element $x_1 \wedge x_2 \wedge x_3 \in S^c(X_\bullet)$. In that case,
\begin{align}
\Delta(x_1 \wedge x_2 \wedge x_3) &= (x_1 \wedge x_2) \otimes x_3 + (-)^{x_2 x_3}(x_1 \wedge x_3) \otimes x_2 + (-)^{x_1(x_2 + x_3)} (x_2 \wedge x_3)\otimes x_1 \\
& + x_1 \otimes (x_2 \wedge x_3) + (-)^{x_1 x_2} x_2 \otimes (x_1 \wedge x_3) + (-)^{x_3(x_1 + x_2)} x_3 \otimes (x_1 \wedge x_2).
\end{align}
The signs in front of each term are just obtained by permuting the three elements to a particular order. The definition of the coproduct to higher powers continues along the same lines. The general formula reads
\begin{equation}
\Delta(x_1 \wedge \cdots \wedge x_n) = \sum_{l = 1}^{n-1} \sum_{\sigma \in S(l,n-l)} \pm (x_{\sigma(1)} \wedge \cdots \wedge x_{\sigma(l)}) \otimes (x_{\sigma(l+1)} \wedge \cdots \wedge x_{\sigma(n)}).
\end{equation}
Here, the second sum runs over a subset $S(p,q)$ of the permutation group of $p+q$ elements. An element $\sigma$ in there is called a \emph{(p,q)-unshuffle}. It is a permutation  with the condition that $\sigma(i) \le \sigma(j)$ whenever $1 \le i \le j \le p$ or $p+1 \le i \le j \le p+q$. The permutation $\sigma$ then determines the sign, which is just obtained by a graded permutation of the elements by $\sigma$.

We now establish some properties of this coalgebra. 
The coproduct is coassociative, which means that
\begin{equation}
(\Delta \otimes \id) \circ \Delta = (\id \otimes \Delta) \circ \Delta.
\end{equation}
This property allows one to uniquely define higher coproducts
\begin{equation}
\Delta_n := (\Delta_{n-1}\otimes \id) \circ \Delta = (\id \otimes \Delta_{n-1}) \circ \Delta, \quad \Delta_1 := \id.
\end{equation}
We now turn to the definition of cohomomorphisms, where the higher coproducts will be useful. 
A cohomomorphism $\phi: S^c(X_\bullet) \rightarrow S^c(Y_\bullet)$ between symmetric tensor coalgebras is a map of degree zero that preserves the coproduct. This means that
\begin{equation}
(\phi \otimes \phi) \circ \Delta = \Delta \circ \phi.
\end{equation}
A useful fact is that such a cohomomorphism is uniquely determined by its image in $Y_\bullet \subseteq S^c(Y_\bullet)$. In other words, defining $\pi_1: S^c(Y_\bullet) \rightarrow Y_\bullet$ as the projection onto $Y_\bullet$, the association $\phi \mapsto \pi_1 \circ \phi$ is bijective as a map from cohomomorphisms to linear maps of degree zero. The inverse is given by
\begin{equation}
f \mapsto \exp(f) := \sum_{n \ge 1} \frac{1}{n!} f^{\wedge n} \circ \Delta_n\,,
\end{equation}
where $f^{\wedge n}: S^c(X_\bullet)^{\otimes n} \rightarrow S^c(Y_\bullet)$ is given by
\begin{equation}
f^{\wedge n}(l_1 \otimes \cdots \otimes l_n) := f(l_1) \wedge \cdots \wedge f(l_n).
\end{equation}

Homomorphisms of coalgebras may seem very strange to readers not familiar with the topic. A more familiar way to think about cohomomorphisms is as non-linear maps between vector spaces. This can be found for example in \cite{markl2015nc}. On the one hand, given a map $f: S^c(X_\bullet) \rightarrow Y_\bullet$, we can construct a formal Taylor series
\begin{equation}\label{Taylor}
f(x) := \sum_{n \ge 1} \frac{1}{n!} f_n(x,...,x)\,,
\end{equation}
where $f_n$ is the $n$-linear piece of $f$. We say that this is formal, since we do not care about convergence of this Taylor series. On the other hand, we can always obtain a cohomomorphism from a formal Taylor series by reading off the $n$-linear piece. Associating a cohomomorphism to a formal Taylor series is an isomorphism in the sense that composing cohomorphisms is the same as composing the Taylor series. In other words, given two cohomomorphisms $\phi_1: S^c(X_\bullet) \rightarrow S^c(Y_\bullet)$ and $\phi_2: S^c(Y_\bullet) \rightarrow S^c(Z_\bullet)$, we have that
\begin{equation}
(\pi_1 \circ \phi_2 \circ \phi_1)(x) = \pi_1 \circ \phi_2(\pi_1 \circ \phi_1 (x)).
\end{equation}
This ensures that the two pictures are indeed equivalent.

The last fact about cohomomorphisms we need for later applications  is that a cohomorphism $f: S^c(X_\bullet) \rightarrow Y_\bullet$ is invertible if and only if its linear piece is. We do not give a full proof of this fact, but only the basic argument. It is based on the Taylor series picture. We write
\begin{equation}
y = f(x) = f_1(x) + \sum_{n \ge 2} \frac{1}{n!} f_n(x,...,x).
\end{equation}
We want to solve for $y$ perturbatively. Assuming that $f_1$ is invertible, we can write
\begin{equation}
x = f_1^{-1}(y) -  \sum_{n \ge 2} \frac{1}{n!} f_1^{-1}\circ f_n(x,...,x).
\end{equation}
We can now plug in the left hand side into the right hand side over and over again. Notice that the $p+1$-linear term in $y$ no longer changes after the $p$th iteration of this process. In this way, we are able to perturbatively determine the $p$-linear piece of the inverse to any order.

Having discussed cohomomorphisms, we now turn to coderivations. A linear map $Q: S^c(X_\bullet) \rightarrow S^c(X_\bullet)$ is called a coderivation, if it satisfies the co-Leibniz rule
\begin{equation}
\Delta \circ Q = (Q \otimes \id + \id \otimes Q) \circ \Delta.
\end{equation}
Similar to cohomomorphisms, a coderivation is determined by its image in $X^\bullet \subseteq S^c(X^\bullet)$. The lift is of course different in this case. Let $b_k: X_\bullet^{\wedge k} \rightarrow X_\bullet$ be an $k$-linear map. We obtain a coderivation by defining
\begin{equation}
b_k(x_1 \wedge \cdots \wedge x_n) := \sum_{\sigma \in S(k,n-k)} \pm b_k(x_{\sigma(1)},...,x_{\sigma(k)}) \wedge x_{\sigma(k+1)} \wedge \cdots \wedge x_{\sigma(n)}.
\end{equation}
The sign is again obtained by permuting the graded vectors with the permutation $\sigma$. Also, the above is defined to be zero when $k > n$. From this perspective, the data of a coderivation is equivalent to a collection of multilinear maps $\{b_k\}_{k \ge 1}$.

\subsubsection*{$L_\infty$-algebras}

We now have gathered all the tools to define an $L_\infty$-algebra. Given a graded vector space $X_\bullet$, an $L_\infty$-algebra is given by a coderivation $Q$ of degree $-1$ on $S^c(X_\bullet)$, such that $Q^2 = 0$. A morphism of $L_\infty$-algebras $\phi: (X_\bullet,Q_1)\rightarrow (Y_\bullet, Q_2)$ is a cohomomorphism, such that $\phi \circ Q_1 = Q_2 \circ \phi$. The simplicity of the square zero condition and the definition of morphism is the main reason to use the language of coalgebras.

In the previous section we explained  that a coderivation is equivalent to a collection of graded symmetric maps $b_k: X^{\wedge k}_\bullet \rightarrow X_\bullet$. The condition $Q^2 = 0$ defines relations among  the $b_k$. There are infinitely many of them, and we explicitly state the first few. In order to do so, let us first  note the following. Coderivations, just as derivations, form a graded Lie algebra with respect to the commutator bracket $[Q_1,Q_2] := Q_1 \circ Q_2 - (-)^{Q_1 Q_2} Q_2 \circ Q_1$. This in particular means that $Q^2 = \frac{1}{2}[Q,Q]$ is a coderivation, independent of whether it is zero or not. This implies in particular that it is identically zero, if and only if it is zero when its output is projected to $X_\bullet$. Hence, all relations can be stated in terms of projecting to one output. One finds the following conditions
\begin{align}
0 =& \ b_1^2, \label{c1}\\
0 =& \ b_1 (b_2(x,y)) + b_2(b_1(x),y) + (-)^x b_2(x,b_1(y)), \label{c2}\\
0 =& \ b_1 b_3(x,y,z) + b_3(b_1(x),y,z) + (-)^x b_3(x,b_1(y),z) + (-)^{x + y} b_3(x,y,b_1(z)), \nonumber \\
&+ b_2(b_2(x,y),z) + (-)^{z}b_2(b_2(z,x),y) + (-)^{y+z}b_2(b_2(y,z),x), \label{c3}\\
\vdots \nonumber
\end{align}
Equation (\ref{c1}) tells us that $(X_\bullet,b_1)$ defines a chain complex. Condition (\ref{c2}) means that the differential $b_1$ acts as a derivation with respect to the product $b_2$. Finally, (\ref{c3}) says that $b_2$ satisfies some Jacobi identity, up to terms that involve $b_1$ and $b_3$.

Let us look at some examples. We first consider  a differential graded Lie algebra, which is a very special type of $L_\infty$-algebra. In that case, all $b_n$ are zero for $n \ge 3$. They occur for example in the theory of Lie algebra valued forms. Given some manifold $M$ and a Lie algebra $\mathfrak{g}$, we can consider the complex $\Omega^\bullet(M) \otimes \mathfrak{g}$. We have a linear map $\tilde{b}_1 = \text{d}$ given by the de Rham differential and a product $\tilde{b}_2 = [-,-]$ from the Lie bracket. With the usual conventions, the de Rham differential is of degree one and the Lie bracket is graded anti-symmetric of degree zero. To match our conventions, we define a complex $X_\bullet$ such that
\begin{equation}
X_n = \Omega^{1-n}(M) \otimes \mathfrak{g}.
\end{equation}
In this convention, both $\tilde{b}_1$ and $\tilde{b}_2$ have degree $-1$. In order to match the signs in the conditions (\ref{c1} - \ref{c3}), we set
\begin{equation}
b_1(x) = \tilde{b}_1(x), \quad b_2(x,y) = (-)^x \tilde{b}_2(x,y).
\end{equation}

The above example shows how $L_\infty$-algebras generalize ordinary Lie algebras. In the Lie algebra example, the Jacobi identity in (\ref{c3}) is strictly satisfied, so there is no need to have a non-trivial $b_3$. On the other hand, whenever there is a $b_3$, one occasionally says that $b_2$ satisfies the Jacobi identity up to homotopy $b_3$. This is one reason why $L_\infty$-algebras are also known as homotopy Lie algebras.

In physics, $L_\infty$-algebras arise as a formulation of classical field theories. We do not go into detail in how this connection is established precisely, but merely state the idea. In the case of a simple gauge theory, one often introduces four vector spaces ranging from degree one to minus two. In the previous sections, we already saw that the free theory is then described by a chain complex
\begin{equation}
\begin{tikzcd}
0 \arrow[r] & X_1 \arrow[r,"\p_1"] & X_0 \arrow[r,"\p_0"] & X_{-1} \arrow[r,"\p_{-1}"] & X_{-2} \arrow[r] & 0.
\end{tikzcd}
\end{equation}
Here, $X_0$ is the space of fields. To the left we have the space $X_1$ of gauge parameters. Given $\Lambda \in X_1$ and $\phi \in X_0$, the differential $b_1$ defines a gauge transformation $\d_\Lambda \phi = \p_1 \Lambda$. The free equations of motion are encoded in the condition $\partial_0 \phi = 0$. Gauge independence of the equations of motion reads $\partial_0 \circ \partial_1 = 0$. The physical content of the theory is therefore encoded in the homology $H_0(X_\bullet) = \frac{\ker \p_0}{\text{im} \, \p_1}$. Since $\partial_0 \phi$ gives the equation of motion, we think of $X_{-1}$ as the space of equations of motion. Finally, we call $X_{-2}$ the space of generalized Bianchi identities. The Bianchi 
identities are encoded in $\partial_{-1} \circ \partial_0 = 0$.
The complex can continue to other degrees if the theory has reducible gauge symmetries. Higher degrees then encode gauge symmetries among gauge symmetries, and gauge symmetries among gauge symmetries of gauge symmetries, etc. This in turn leads to higher Bianchi identities in negative degrees.

To capture interactions in a theory, we need to extend the data of a chain complex to that of an $L_\infty$-algebra. Given a field $\phi \in X_0$, we would like to write the non-linear equations of motion in a power series as
\begin{equation}\label{MCequation}
0 = \sum_{n \ge 1} \frac{1}{n!}b_n(\phi,...,\phi).
\end{equation}
In this context, the equations of motion are also known as Maurer-Cartan equations. On the other hand, gauge symmetries in an interacting theory may be non-linear. Given a gauge parameter $\Lambda \in X_1$, we write the gauge transformation it generates on a field $\phi$ as
\begin{equation}\label{generaltransformation}
\delta_\Lambda \phi = \sum_{n \ge 1}\frac{1}{(n-1)!} b_n(\phi,...,\phi,\Lambda).
\end{equation}
Here, the $L_\infty$-relations become important. The theory is gauge invariant precisely when the relations are satisfied. For example, in Chern-Simons theory one precisely uses the fact that we have a Lie bracket in order to prove  gauge invariance. In other degrees, there are also relations about non-linear Bianchi identities and higher gauge structure.

\subsubsection*{$L_\infty$-morphisms and gauge transformations}

We now aim to explain in a little more detail how to translate from the coalgebra language to more  familiar formulas for the gauge transformations. 
We first recall that according to  \eqref{Taylor} any coalgebra morphism $f: S^c(X_\bullet) \rightarrow Y_\bullet$ defines a formal Taylor series
\begin{equation}
f(x) = \sum_{n \ge 1} \frac{1}{n!} f_n(x,...,x).
\end{equation}
We defined $L_\infty$-morphisms $f: (X_\bullet,Q_1) \rightarrow (Y_\bullet, Q_2)$ to be coalgebra morphisms that commute with the coderivations, i.e.~$Q_2 \circ f = f \circ Q_1$. 
We will now show that $f$, when viewed  as a function, is covariant under gauge transformations. 
We need to introduce some conventions. For now, we denote by $b := \pi_1 \circ Q_1$ the part of $Q_1$ with one output. Also, %given an element $x$ of degree zero, 
we introduce the exponential as the map $X_{0}\rightarrow S^c(X_{\bullet})$,  defined for an element $x$ of degree zero by  
\begin{equation}
\exp(x) := \sum_{n \ge 1} \frac{1}{n!} x^{\wedge n}\,. 
\end{equation}
This map  has the characteristic property 
 \be
  \Delta(\exp(x))=\exp(x)\otimes \exp(x)\,. 
 \ee
From this it follows that for a coalgebra morphism $f$ we have 
\begin{equation}\label{fandexpcommute}
f(\exp(x)) = \exp (f(x)).
\end{equation}
By a slight abuse of notation, on the left-hand side we view $f$ as  a coalgebra morphism
and on the right-hand side we view $f$ as a map $X_{\bullet}\rightarrow Y_{\bullet}$. 
We can now write the gauge transformation (\ref{generaltransformation}) as
\begin{equation}\label{GT1}
\delta_\Lambda \phi %= b(\exp(\phi)\Lambda) 
= \frac{\partial}{\partial \varepsilon} b(\exp(\phi+ \varepsilon \Lambda)),
\end{equation}
where we introduced a formal parameter $\varepsilon$ of degree minus one, and  we take the odd $\varepsilon$ to be nilpotent, $\varepsilon^2=0$.

For the following to work, we need to assume that the part of $f$ that takes values in the field space $Y_0$ only depends on fields in $X_0$. In this case, we can effectively assume that
\begin{equation}\label{assumption}
b(\exp(\phi + \varepsilon\Lambda)) = b(\exp(\phi + \varepsilon\Lambda))\exp(\phi + \varepsilon\Lambda) \simeq b(\exp(\phi + \varepsilon\Lambda))\exp(\phi)\,,
\end{equation}
which holds as arguments of $f$, and 
 we used $\simeq$ instead of the equality sign  to indicate that the above assumption on $f$ is needed. We now apply the $L_{\infty}$-morphism $f$ to both sides of \eqref{GT1} and use \eqref{fandexpcommute} as well as \eqref{assumption}. We find 
\begin{equation}\label{Covariancechainmaps}
\begin{split}f(\delta_\Lambda \phi) &= f(\delta_\Lambda\exp(\phi)) = \frac{\p}{\partial \varepsilon}f(b(\exp(\phi + \varepsilon\Lambda))\exp(\phi)) = \frac{\p}{\partial \varepsilon} f(b(\exp(\phi + \varepsilon\Lambda)))  \\
&= \frac{\p}{\partial \varepsilon} b(\exp(f(\phi+\varepsilon \Lambda))) = \frac{\p}{\partial \varepsilon} b(\exp(f(\phi)+\varepsilon f^{(1)}(\phi,\Lambda))) =  \delta_{f^{(1)}(\phi,\Lambda)} f(\phi) \,,
\end{split}
\end{equation}
where we used 
 \be
   f(\phi+\varepsilon \Lambda)=f(\phi)+\varepsilon f^{(1)}(\phi,\Lambda)\,, 
 \ee
with 
\begin{equation}
f^{(1)}(\phi,\Lambda) = \sum_{n \ge 1} \frac{1}{(n-1)!} f_n(\phi,...,\phi,\Lambda).
\end{equation}
Equation \eqref{Covariancechainmaps} shows that $L_\infty$-morphisms, when thought of as functions, are covariant with respect to gauge transformations. In particular, if the target space has no gauge transformations, $f$ is gauge invariant.

One may wonder what happens when we do not make the assumption on $f$. In that case additional terms proportional to the equations of motion enter. In that case, $f$ is only gauge invariant on shell (assuming that $Y_\bullet$ has no gauge transformations, i.e. $Y_1 = 0$). One would obtain such $f$ when applying the procedure explained below to theories with open gauge algebras.

\subsection{Homotopy transfer and the homological perturbation lemma}

A central concept of $L_\infty$-algebras is that of homotopy transfer, see \cite{vallette2014algebra}.\footnote{We note that homotopy transfer 
is also closely related to the elimination of generalized auxiliary fields in the BV-BRST formulation of gauge theories, see \cite{Henneaux:1989ua,Barnich:2004cr}.} One asks the following question: 
Suppose that we have a pair of chain complexes $(X_\bullet,b_1)$ and $(Y_\bullet,c_1)$, and suppose that 
$X_\bullet$ is equipped with an $L_\infty$-structure, given by an infinite tower of products $\{b_i\}_{i \ge 1}$ and such that $b_1$ 
is the differential of the chain complex $X_\bullet$, 
under which conditions can we then transfer the $L_\infty$-structure of $X_\bullet$ to $Y_\bullet$, such that the two structures are equivalent?
This is possible provided $Y_\bullet$ is a strong deformation retract of $X_\bullet$,\footnote{A strong deformation retract is actually not required, but it will be necessary for applying the homological perturbation lemma introduced below.} which means that there are chain maps $p_0: X_\bullet \rightarrow Y_\bullet$ and ${\iota}_0: Y_\bullet \rightarrow X_\bullet$, together with a degree one map $s: X_\bullet \rightarrow X_\bullet$, such that
\begin{equation}
1 - {\iota}_0\circ p_0 = b_1 s + s b_1, \quad p_0 \circ {\iota}_0 = 1, \quad s \circ {\iota}_0 = 0, \quad p_0 \circ s = 0, \quad s^2 = 0.
\end{equation}
Here we use the subscript 0 to distinguish these maps from deformations to be provided by the deformation lemma.  
The first of the above conditions implies that ${\iota}_0$ and $p_0$ are inverse on homology. 
Then there exists an $L_\infty$-structure $\{c_n\}_{n \ge 1}$ on $Y_\bullet$, together with $L_\infty$-morphisms $p: (X_\bullet,\{b_n\}_{n \ge 1}) \rightarrow (Y_\bullet,\{c_n\}_{n \ge 1})$ and ${\iota}: (Y_\bullet,\{c_n\}_{n \ge 1}) \rightarrow (X_\bullet,\{b_n\}_{n \ge 1})$ such that $p \circ {\iota} = 1$, and their linear pieces are given by $p_0$ and ${\iota}_0$, respectively. More generally, we say that two $L_\infty$-structures are equivalent, if there exists a morphism such that its linear part is an isomorphism on homology. So the $L_\infty$-structures on $X_\bullet$ and $Y_\bullet$ are equivalent, precisely since ${\iota}_0$ (or $p_0$) is an isomorphism on homology.

The formulas for the morphisms and the induced $L_\infty$-structure are given perturbatively as cohomomorphisms and coderivation respectively.  The fact that $p_0$ and ${\iota}_0$ are linear simplifies their lift drastically. We have
\begin{equation}
p_0(x_1\wedge \cdots \wedge x_n) = p_0(x_1)\wedge \cdots \wedge p_0(x_n), \quad {\iota}_0(y_1\wedge \cdots \wedge y_n) = {\iota}_0(y_1)\wedge \cdots \wedge {\iota}_0(y_n).
\end{equation}
We also have to construct a lift of the homotopy $s$ to coalgebras. We first define it on tensor powers
\begin{align}
s_n(x_1 \otimes \cdots \otimes x_n) =\,& s(x_1) \wedge x_2 \cdots \wedge x_n + (-)^{x_1}{\iota}\circ p(x_1) \wedge s(x_2)  \wedge x_3 \cdots \wedge x_n  \\
& + ... + (-)^{x_1 + ... x_{n-1}} {\iota}\circ p(x_1) \wedge \cdots \wedge {\iota} \circ p (x_{n-1}) \wedge s(x_n).
\end{align}
We also define
\begin{equation}
q_n(x_1 \wedge \cdots \wedge x_n) = \frac{1}{n!}\sum_{\sigma \in S_n} \pm x_{\sigma(1)}\otimes \cdots \otimes x_{\sigma(n)},
\end{equation}
where the sum runs over all permutations $\sigma$ of $n$ elements, and the sign is the natural one coming from commuting graded objects. We can now define a map $s_n \circ q_n: X_\bullet^{\wedge n} \rightarrow X^{\wedge n}_\bullet$. Finally, we obtain the map $s := \sum_{n \ge 1} s_n \circ q_n$ from the coalgebra $S^c(X_\bullet)$ to itself.

The perturbed $L_\infty$-morphisms, as well as the induced products, can be obtained by applying the homological perturbation lemma, see \cite{crainic2004perturbation} 
for a nice discussion. They read
\begin{equation}
{\iota} = (1 + s \circ \delta)^{-1}\circ {\iota}_0, \quad p = p_0 \circ (1 + \d \circ s)^{-1}, \quad Q = c_1 + \delta \circ (1 + s \circ \d)^{-1}.
\end{equation}
Here, $\delta := \sum_{n \ge 2} b_n$ is the coderivation corresponding to the products defining the $L_\infty$-structure on $X_\bullet$, but without the linear piece. Also, we define the inverses appearing in the above formulas as a geometric series, i.e.~$(1 + x)^{-1} = \sum_{n \ge 0} (-x)^n$. The fact that ${\iota}$ and $p$ are coalgebra morphisms is guaranteed by the fact that we considered $s$ to be a strong deformation retract \cite{huebschmann1991 ,huebschmann2011sh}.

Let us recapture the essence of the homotopy transfer theorem from the point of view of field theory. We noted earlier that free field theories are equivalent if the associated complexes are homotopic to each other (in particular, the fields are equivalent on-shell and up to gauge transformations). In general, such a homotopy arises either by gauge fixing some of the fields or by solving equations of motion. One then obtains an effective theory in terms of a smaller set of fields. The homotopy transfer then allows us to carry this over to an interacting theory. Applying the procedure allows us to extract from any field theory an effective theory of a smaller set of fields. The projection map $p: (X_\bullet, \{b_n\}_{n \ge 1}) \rightarrow (Y_\bullet, \{c_n\}_{n \ge 1})$ gives rise to a non-linear relation between the old (larger) set of fields in $X_\bullet$ and the new smaller one in $Y_\bullet$. The fact that $p$ is an $L_\infty$-morphism ensures that the relation is covariant with respect to gauge transformations and that equations of motion are correctly translated from the original theory to the effective theory. More precisely, if $x$ solves the Maurer-Cartan equation on $X_\bullet$, then $p(x)$, where $p(x)$ is given by a formal Taylor series as in (\ref{Taylor}), solves the Maurer-Cartan equation on $Y_\bullet$. The equivalence property (i.e.~that the linear part of $p$ is an isomorphism on homology) ensures that sets of solutions modulo gauge transformations of both theories are actually isomorphic \cite{Fukaya}.

Below we will use the homotopy transfer as a tool to remove gauge degrees of freedom. The main advantage of this approach is that the map $p$ produced by the transfer theorem automatically gives a relation between the original gauge redundant fields, and the new gauge invariant  fields. Moreover, the fact that $p$ is an $L_\infty$-morphism implies that the relation is invariant under the full non-linear gauge transformations  of the original space. 
We define the gauge invariant variable by $\widehat{\phi}\equiv p(\phi)$. Its gauge invariance follows  from \eqref{Covariancechainmaps}: 
\begin{equation}\label{invariatproj}
\delta_{\Lambda}\widehat{\phi}=p(\delta_\Lambda \phi) = \delta_{p^{(1)}(\phi,\Lambda)} p(\phi)=0 \, , 
\end{equation}
which vanishes since $p^{(1)}(\phi,\Lambda) \in Y_1 = 0$ for the complex of gauge invariant variables. Therefore, $\widehat{\phi}= p(\phi)$ is gauge invariant.

\subsection{Yang-Mills theory}

\subsubsection*{$L_\infty$-algebra description}

The free theory of Yang-Mills has already been covered in sec.~2. We now give the description of the full interacting theory as an $L_\infty$-algebra, beginning 
with the gauge transformations. The non-linear gauge transformation of Yang-Mills theory reads
\begin{equation}
\delta_\Lambda A_\mu = \partial_\mu \Lambda + [A_\mu,\Lambda].
\end{equation}
When we compare this to (\ref{generaltransformation}), we find that we have to set
\begin{equation}
b_2^\mu(A, \Lambda) = b_2(\Lambda,A^\mu) = [A^\mu,\Lambda].
\end{equation}
There are no higher products as long as a gauge parameter is involved, i.e. $b_k(\Lambda,...) = 0$ for all $k \ge 3$. The equations of motion read
\begin{equation}
0 = \Box A_\mu - \partial_\mu \p \cdot A + [A^\nu, \partial_\nu A_\mu - \p_\mu A_\nu] + \p^\nu [A_\nu,A_\mu] + [A^\nu,[A_\nu,A_\mu]].
\end{equation}
When we compare this to the general form (\ref{MCequation}), we find that,
\begin{equation}
b_2^\mu(A,B) = [A_\nu,\partial^\nu B^\mu - \partial^\mu B^\nu] + (A \leftrightarrow B)
\end{equation}
and
\begin{equation}
b_3^\mu(A,B,C) = [A_\nu,[B^\nu,C^\mu]] + \text{permutations},
\end{equation}
where $A_\mu,B_\nu,C_\rho$ are gauge fields. Higher brackets are zero.

The above definitions correctly reproduce gauge transformations and equations of motion. However, we also want to include the space $X_{-2}$ of Bianchi identities. Moreover, even without it, one can check that the above products do not satisfy the $L_\infty$-relations. There are two reasons. At the non-linear level, gauge transformations are not abelian, but are again proportional to a gauge transformation. We can define
\begin{equation}
b_2(\Lambda_1,\Lambda_2) = - b_2(\Lambda_2,\Lambda_1) = -[\Lambda_1,\Lambda_2].
\end{equation}
Another issue is that a gauge transformation of the equations of motion does not give zero, but is again proportional to the equations of motion. The equation of motion $D_\nu F^{\nu\mu}$ transforms in the adjoint representation of the gauge algebra. To account for this fact, we should define a product between gauge parameters $\Lambda \in X_1$ and equations of motion $E^\mu \in X_{-1}$. We set
\begin{equation}
b_2^\mu(E,\Lambda) = - b_2^\mu(\Lambda,E) = [E^\mu,\Lambda].
\end{equation}

To obtain the correct Bianchi identities, we need to introduce brackets between scalars $B \in X_{-2}$ and gauge parameters. By degree reasons, this is the only non-vanishing combination. We can determine the bracket via the rule
\begin{equation}
\p b_2(E,\Lambda) = -b_2( \p E, \Lambda) - b_2(E,\Lambda).
\end{equation}
The left hand side is
\begin{equation}
\partial_\mu [E^\mu,\Lambda] = [\partial_\mu E^\mu, \Lambda] + [E^\mu,\partial_\mu \Lambda].
\end{equation}
From this it follows that
\begin{equation}
b_2(B,\Lambda) = m_2(\Lambda,B) = -[B,\Lambda], \qquad b_2(E,A) = b_2(A,E) = -[E^\mu,A_\mu],
\end{equation}
where $\Lambda \in X_1, A_\mu \in X_0, E^\mu \in X_{-1}, B \in X_{-2}$.

\subsubsection*{Homotopy transfer}

We recall the homotopy from the complex $X_\bullet$ to the gauge invariant complex $\bar{X}_\bullet$. We denote this by 
\begin{equation}
\begin{tikzcd}
0 \arrow[r] & X_1 \arrow[r,"\p"] \arrow[d,"p_0"] & X_0 \arrow[r,"\p"] \arrow[d,"p_0"] & X_{-1} \arrow[r,"\p"] \arrow[d,"p_0"] & X_{-2} \arrow[d,"p_0"] \arrow[r] & 0 \\
0 \arrow[r] & 0 \arrow[r] & \bar{X}_0 \arrow[r,"\bar{\partial}"] & \bar{X}_{-1} \arrow[r] & 0 \arrow[r] & 0
\end{tikzcd} \ .
\end{equation}
Since $p_0 {\iota}_0 = 1$ and therefore $\bar{\partial} = \bar{\partial} p_0 {\iota}_0 = p_0 \partial {\iota}_0$, $\bar{\partial}$ is obtained by restricting $\partial$ to $\bar{X}_\bullet$. On gauge fields $A_\mu$, $s$ is given by
\begin{equation}
s(A) = \Delta^{-1}\partial_i A^i.
\end{equation}
This allows us to write
\begin{equation}
A_\mu = \bar{A}_\mu + \partial_\mu s(A),
\end{equation}
where $\bar{A}_\mu$, as a function of $A_\mu$, is invariant under gauge transformations. In the notation of section 2, we would write
\begin{equation}
\bar{A}_\mu = p_0(A_\mu) = A_\mu - \partial_\mu \Delta^{-1}\partial^i A_i .
\end{equation}

We can now apply the homotopy transfer to obtain a relation
\begin{equation}
\widehat{A}_\mu = p(A_\mu)
\end{equation}
that is invariant under non-linear gauge transformations. Recall that $p$ was given by
\begin{equation}
p = p_0 (1 + \delta s)^{-1},
\end{equation}
where $\delta = \sum_{k \ge 2} b_k$. All maps involved are lifted so that they act on coalgebras. To get an idea what $p$ looks like, we expand it to quadratic order. Its linear part is just $p_0$, so we directly turn to its quadratic part. Given two gauge fields $A,B \in X_0$, it is
\begin{align}
- p_0 b_2 s (A \wedge B) &= - p_0 b_2 \frac{1}{2} (s(A) \wedge B + A \wedge s(B) + s(A) \wedge {\iota}_0p_0(B) + {\iota}_0p_0(A) \wedge s(B)) \\
&= - \frac{1}{2} p_0  ([(1 + {\iota}_0p_0)(A_\mu),s(B)] - [s(A),(1 + {\iota}_0 p_0)(B)]).
\end{align}
We now turn this into a non-linear relation between $\widehat{A}_\mu$ and $A_\mu$. In order to do this, we need to insert the appropriate symmetry factors.
\begin{align}
\widehat{A}_{\mu}= p(A_\mu) &= p_0(A_\mu) - \frac{1}{4} p_0([(1 + {\iota} p)(A_\mu),s(A)] - [s(A), (1 + {\iota}p)(A_\mu)]) \\
&=  p_0(A_\mu) + \frac{1}{2} p_0 [s(A), (1 + {\iota}p)(A_\mu)]\,, \label{gito2ndorder}
\end{align}
which is the relation (\ref{IntrINVREL}) quoted in the introduction.

As a consistency check, we now want to proof that $\widehat{A}_\mu$ is indeed gauge invariant to second order. We first compute the linear variation of the quadratic term. Using $s(\partial_\mu \Lambda) = \Lambda, \partial_\mu s(A_\mu) = (1 - {\iota}p)(A), p(\partial_\mu ...) = 0$, we find that
\begin{align}
\frac{1}{2} p_0 [s(\partial_\mu \Lambda), (1 + {\iota}p)(A_\mu)] + \frac{1}{2} p_0 [s(A), (1 + {\iota}p)(\partial_\mu \Lambda)] \\
= \frac{1}{2} p_0 ([\Lambda, (1 + {\iota}p)(A_\mu)] + [s(A),\partial_\mu \Lambda]) \\
= \frac{1}{2} p_0 ( [\Lambda, (1 + {\iota}p)(A_\mu)] - [(1 - {\iota}p)(A_\mu), \Lambda]) \\
= - p_0([A_\mu,\Lambda]). \label{QaudraticVariation}
\end{align}
On the other hand, the variation of the linear part reads
\begin{equation}
p_0(\partial_\mu \Lambda + [A_\mu,\Lambda]) = p_0([A_\mu,\Lambda]),
\end{equation}
which exactly cancels \eqref{QaudraticVariation}. This proves gauge invariance of $\widehat{A}^\mu$ to second order.

We now come to the question what the effective equations of motion are. Recall that the general formula for the associated coderivation is
\begin{equation}
\bar{Q} = \bar{\partial} + p_0 \delta (1+s \delta)^{-1} {\iota}_0.
\end{equation}
 $\bar{Q}$ simplifies drastically in our case. First of all, since $\bar{X}_\bullet$ only contains fields in degree zero and one, $\bar{Q}$ can only be nonzero when acting on degree zero elements, i.e.~on gauge fields $\widehat{A}_\mu$ only. Next we note that
\begin{equation}\label{sum}
(1 + s\delta)^{-1} = \sum_{n \ge 0} (-s\delta)^n
\end{equation}
 acts as the identity on these fields. The reason is that $\delta$, when restricted to fields in $X_0$, produces an element in $X_{-1}$ . But $s$ is defined to be zero on these elements. The sum in \eqref{sum} breaks after the zeroth term. This implies that
\begin{equation}
\bar{Q} = \bar{\partial} + p_0\delta i_0 = p_0(\partial + \delta){\iota}_0.
\end{equation}
Therefore, the full equations of motion can be obtained by restriction to $\partial_i A^i = 0$.

\subsubsection*{Gauge invariant action}

We now want to explain why  the equations of motion (and the action) can be obtained by simply replacing  $A_\mu$ by $\widehat{A}_\mu$. 
We will argue, within the $L_{\infty}$ framework, that there is a  map $\phi(A)$ to the space of gauge parameters so that together 
with the map $\widehat{A}(A)$ constructed above the following  identity, c.f.~(\ref{NewIdentity}) in the introduction, holds: 
  \be\label{NewIdentity2}
  A_{\mu} \ = \ e^{\Delta_{\phi(A)}}\,\widehat{A}_{\mu}\;. 
 \ee 
Here $\Delta_\phi$ is the operator defining  infinitesimal gauge transformations, $\Delta_\phi(A_\mu) = \partial_\mu \phi + \big[A_\mu,\phi\big]$.  
 Since this takes the form of a finite gauge transformation, it follows from gauge invariance of the action  that we may simply replace 
 $A_{\mu}$ by $\widehat{A}_{\mu}$.  
 
 In the last section we constructed a homotopy from $X_\bullet$ to $\bar{X}_\bullet$. The homotopy forgets about some of the physically irrelevant data (in this case the gauge degrees of freedom). We now find it convenient  to still pass from $A_\mu$ to $\bar{A}_\mu$, but to not forget about the gauge degrees of freedom. 
Rather, we represent the gauge degrees of freedom by  another field $\phi$, denoting  the space encoding $\phi$ by $Y_0$. 
Let us  now define the map 
\be\label{calFDEF}
\begin{split}
F: \,\bar{X}_0 \oplus Y_0 	&\rightarrow X_0, \\
(\bar{A}_\mu,\phi)		&\mapsto {\iota}_0(\bar{A}_\mu) + j_0(\bar{A}_\mu,\phi),
\end{split}
\ee
where   
$j_0(\bar{A}_\mu,\phi) = \partial_\mu \phi$. 
The inverse map is given by 
 \be\label{DefInverseF}
 \begin{split}
  F^{-1}: X_0& \rightarrow \bar{X}_0\oplus Y_0\\
  A_\mu & \mapsto (p_0(A_\mu),q_0(A_\mu))\,,
 \end{split} 
 \ee 
where $q_0$ is defined like the homotopy, $q_0=\Delta^{-1}\partial_iA^i$, but viewed as a degree 0 map. 
With this it is straightforward to verify that these maps are indeed inverses to each other. 
These maps formalize the intuitively clear fact that, at least at the linearized level, 
the space of all gauge fields $X_0$ is isomorphic (of the same `size') to the sum of  the space of gauge invariant 
fields plus the space of pure gauge fields. 

Our goal is now to extend the above maps to the non-linear or interacting level by constructing non-linear  extensions $q$ and $j$ of $q_0$ and $j_0$. 
This mimics the way the perturbation lemma provides non-linear corrections to $p_0$. 
In view of the desired identity (\ref{NewIdentity2}), we  define the map ${\cal F}: \bar{X}_0 \oplus Y_0 \rightarrow X_0$ given by 
\begin{equation}
{\cal F}(\widehat{A}_\mu,\phi) = e^{\Delta_\phi}\widehat{A}_\mu\;, 
\end{equation}
where $\widehat{A}_\mu\in \bar{X}_0$ is denoted by a hat to indicate that this will be the fully gauge invariant field. 
The map ${\cal F}$ serves as a nonlinear extension of the maps ${\iota}_0$ and $j_0$, since to linear order we have with (\ref{calFDEF}) 
\begin{equation}\label{Psiatlinear}
{\cal F} = {\iota}_0 + j_0 + \cdots  \, .
\end{equation} Recall that, in the $L_\infty$-setup, maps are invertible perturbatively if and only if their linear part is. Because of \eqref{Psiatlinear}, this is true for ${\cal F}$. In this way, we can in principle determine the non-linear extension ${\cal F}^{-1}$ of (\ref{DefInverseF}) and hence the non-linear extension $q$ of $q_0$. 
This in turn determines the function $\phi(A)$ perturbatively from the ansatz 
 \be
  \big(\widehat{A}_{\mu},\phi(A)\big) \equiv  \big(\widehat{A}_{\mu},q(A)\big) \equiv {\cal F}^{-1}(A_{\mu})\;. 
 \ee
 Let us illustrate this to quadratic order, by expanding  ${\cal F}$ to this  order: 
\begin{equation}\label{PertPsi1}
A_\mu = {\cal F}(\widehat{A}_{\mu},\phi) = \widehat{A}_\mu + \partial_\mu \phi + \big[\widehat{A}_\mu,\phi\big] + \frac{1}{2}\big[\partial_\mu \phi,\phi\big] + \cdots  \ .
\end{equation}
We now apply  $q_0$ to both sides. After bringing $A_\mu$ to the right and $\phi$ to the left, and recalling $q_0(\partial_{\mu}\phi)=\phi$ we find
\begin{equation}
\phi = q_0(A_\mu) - q_0\big[\widehat{A}_\mu, \phi\big] - \frac{1}{2}q_0\big[\partial_\mu \phi,\phi\big]\,.
\end{equation}
We then write $\widehat{A}_\mu$ and $\phi$ to linear order in $A_\mu$, recalling (\ref{gito2ndorder}) and that to leading order $\phi(A)=s(A)$, 
\begin{align}
\phi &= q_0(A_\mu) - q_0\big[{\iota}_0p_0(A_\mu),s(A)\big] - \frac{1}{2}q_0 \big[(1- {\iota}_0p_0)(A_\mu),s(A)\big] \nonumber \\
&= q_0(A_\mu) - \frac{1}{2}q_0\big[(1+{\iota}_0p_0)(A_\mu),s(A)\big]\,. \label{phiofA}
\end{align}
This is $q$ to first non-trivial order. It should now be clear that, in principle, we could continue this procedure to determine $\phi$ as a function of $A_\mu$ to arbitrary order.

As a consistency check, we now want to see whether ${\cal F}$ really inverts $A_\mu \mapsto (p(A_\mu),q(A_\mu))$. 
We do this using \eqref{PertPsi1} to express $\widehat{A}_\mu$ as a function of $A_\mu$, and then check whether it matches what we found through the perturbation lemma. Applying $p_0$ to both sides of \eqref{PertPsi1} tells us that
\begin{equation}
\widehat{A}_\mu = p_0(A_\mu) - p_0\big[\widehat{A}_\mu, \phi\big] - \frac{1}{2}p_0\big[\partial_\mu \phi,\phi\big]\,.
\end{equation}
Expressing $\widehat{A}_\mu$ and $\phi$ in terms of $A_\mu$ to linear order gives
\begin{align}
\widehat{A}_\mu 	&= p_0(A_\mu) - p_0\big[{\iota}_0p_0(A_\mu), s(A)\big] - \frac{1}{2}p_0\big[(1-{\iota}_0p_0)(A_\mu),s(A)\big] \nonumber \\
				&= p_0(A_\mu) -\frac{1}{2}p_0\big[(1 + {\iota}_0 p_0)(A_\mu),s(A)\big]\,. \label{AhatofA}
\end{align}
This is exactly what we found in \eqref{gito2ndorder}.

\section{Cosmological Perturbation Theory}

We now turn to the main motivation of this paper, the perturbation theory about cosmological or FLRW backgrounds. 
As an illustration for the  techniques developed in previous sections we determine the general quadratic action of Einstein-Hilbert coupled to a scalar 
expanded around FLRW in terms of gauge invariant variables. We do this both by a brute-force computation and using the methods of 
homotopy transfer, arriving at the same result (that to the best of our knowledge has not appeared before). 

\subsection{Gravity around FLRW}
We consider quadratic perturbations in an FLRW universe with minimally-coupled scalar matter.  The theory is described by the action,
\begin{equation}
\label{Full matter action cosmology}
S \ =\ \int d^4 x\ \sqrt{-g} \ \bigg\{R  - \frac{1}{2} g^{\mu\nu} \partial_\mu \mathcal{X}\partial_\nu \mathcal{X}- V ( \mathcal{X}) \bigg\}\,, 
\end{equation}
where $\mathcal{X}$ denotes the scalar matter field and $V(\mathcal{X})$ is its potential.  The expansion of the fields around the purely time-dependent background is as follows:
\begin{equation}
\begin{split}
 g_{\mu\nu}(\eta, x)  & \ = \ a^2(\eta) ( \eta_{\mu\nu} + h_{\mu\nu} (\eta ,x) )\,,  \\ 
\mathcal{X}(\eta, x) & \ =\ {\mathcal{X}}^{(0)} (\eta)  + \varphi (\eta, x) \,,
 \end{split} 
\end{equation} 
where $\eta$ is conformal time. The background dynamics are governed by the Friedmann equations and the equation of motion of $\mathcal{X}^{(0)}$:
\begin{flalign}
\label{Friedmann1}
&H^2  \ = \ \frac{1}{6}a^2  {\rho}    \,, \\
\label{Friedmann2}
&\dot{H} + H^2   \ = \ \frac{1}{12} a^{ 2}  (   {\rho} - 3  {p} ) \,,\\
\label{eomvarphibar}
&\ddot{\mathcal{X}}^{(0)}  +  2 H    \dot{\mathcal{X}}^{(0)} + a^2   V'\big({\mathcal{X}}^{(0)}\big)  \ = \ 0 \,,
\end{flalign}
where the dot denotes the derivative with respect to conformal time, $H\equiv\frac{\dot{a}}{a}$ is the Hubble parameter, and the prime indicates a derivative with respect to $\mathcal{X}^{(0)}$. ${\rho}$ and ${p}$ are the background density and pressure, respectively:
\begin{flalign}
 {\rho}  \ = \  \frac{1}{2}a^{-2} \dot{\mathcal{X}}^{(0)2}   +  V\big( \mathcal{X}^{(0)} \big)  \,,\\ 
 {p}  \ = \ \frac{1}{2}a^{-2}\dot{\mathcal{X}}^{(0)2}  -    V \big( \mathcal{X}^{(0)} \big)\,,
\end{flalign}
which satisfy the conservation equation
\begin{equation}
\label{conservationbkgrnd}
\dot{ \rho} +  3 H (  {\rho } +  {p} ) =0\,. 
\end{equation}
The gauge transformations of the metric and the scalar matter field are given by the Lie derivative with respect to the gauge parameter $\xi^\mu$:
\begin{equation}
\begin{split}
\delta g_{\mu\nu} & \ =\  \mathcal{L}_{\xi} g_{\mu\nu} \ \equiv \  \xi^\rho\partial_\rho g_{\mu\nu} + \partial_\mu  \xi^\rho g_{\rho\nu} + \partial_\nu \xi^\rho g_{\mu\rho}\;,\\
\delta \mathcal{X}&\ =\  \mathcal{L}_{\xi} \mathcal{X} \ \equiv \ \xi^\rho \partial_\rho \mathcal{X}\,.
\end{split}
\end{equation}
In order to find the gauge transformation of the fluctuation fields, let us take $\xi^\mu$ to be a first-order parameter. We take the  background quantities to be invariant under first-order gauge transformations, and so 
\begin{equation}
\begin{split}
\label{GT g h munu}
\delta g_{\mu\nu}   &\ =\  a^2\delta h_{\mu\nu}\ =\  \xi^\rho\partial_\rho \big[ a^2( \eta_{\mu\nu} + h_{\mu\nu} )\big]  + \partial_\mu  \xi^\rho \big[ a^2( \eta_{\rho\nu} + h_{\rho\nu} )\big] + \partial_\nu \xi^\rho \big[ a^2( \eta_{\mu\rho} + h_{\mu\rho} )\big] \;,\\
\delta \mathcal{X}&\ =\  \delta \varphi \ =\ \xi^\rho \partial_\rho \big( \mathcal{X}^{(0)} + \varphi \big)\;.
\end{split}
\end{equation}
Because both the gauge parameter and the fluctuation are first-order quantities, the second-order terms on the right-hand-sides of (\ref{GT g h munu}) can be neglected. It follows that to lowest order the gauge transformation of the fluctuations are
\begin{equation}
\begin{split}
\delta  h_{\mu\nu} &\ =\ a^{-2}   \xi^\rho\partial_\rho ( a^2 \eta_{\mu\nu} ) + \partial_\mu \xi_\nu + \partial_\nu \xi_\mu \;,\\
\delta \varphi & \ =\ \xi^\rho \partial_\rho \mathcal{X}^{(0)}\;.
\end{split}
\end{equation}
Evaluating this in components,
\begin{equation}
\begin{split} 
\label{PRE Components Transformation COSMO}
\delta h_{00} & \ =\  2 H \xi_0 + 2 \dot{\xi}_0  \,,\\ 
\delta h_{0i} &\  = \   \dot{\xi}_i + \partial_i \xi_0  \,,\\
\delta h_{ij} &\  = \  -2 H \xi_0 \delta_{ij} + 2 \partial_{(i} \xi_{j)}   \,,\\
\delta \varphi &\  = \ -\dot{\mathcal{X}}^{(0)} \xi_0  \,. 
\end{split}
\end{equation}
In order to decompose the scalar, vector, and tensor modes of the fields, we decompose the gauge parameter $\xi_{\mu}$ as prescribed in  (\ref{decompositionxi}). Then the gauge transformations in (\ref{PRE Components Transformation COSMO}) are decomposed as:
\begin{equation}
\begin{split}
\label{COMPONENTS transformation COSMO}
\delta h_{00} & \ =\ 2 H \xi_0 + 2 \dot{\xi}_0  \,,\\ 
\delta h_{0i} & \ =\  \dot{\zeta}_i + \partial_i ( \dot{\chi} + \xi_0) \,,\\
\delta h_{ij} &  \ =\  -2 H \xi_0 \delta_{ij} + 2 \partial_{(i} \zeta_{j)} + 2 \partial_i \partial_j \chi \,,\\
\delta \varphi & \ =\  -\dot{\mathcal{X}}^{(0)} \xi_0  \,. 
\end{split}
\end{equation}
For the metric fluctuations we use the same SVT decomposition as in (\ref{SVTdecompositionTensor}).
By inserting the decomposition into the equations  (\ref{COMPONENTS transformation COSMO}), we find the transformations of the individual components of the metric fluctuations: 
\begin{equation}
\begin{split}
\label{SVT Components Transformation COSMO}
\delta \phi & \ =\  -H  \xi_0 -  \dot{\xi}_0 \,,\\
\delta B_i &  \ =\ \dot{\zeta}_i  \,,\\
\delta B &  \ =\ \dot{\chi} + \xi_0 \,,\\
\delta E_i &  \ =\ \zeta_i \,,\\
\delta E &  \ =\ \chi  \,,\\
\delta C & \ = \- H \xi_0  + \frac{1}{3} \Delta \chi\,.
\end{split}
\end{equation}
By inspecting the transformation of the scalar matter field in (\ref{COMPONENTS transformation COSMO}) and of the metric components in (\ref{SVT Components Transformation COSMO}), one finds (in addition to the gauge invariant tensor $\widehat{h}_{ij}$) the gauge invariant combinations:
\begin{equation}
\begin{split}
\label{BardeenVarsCosmo}
\Sigma_i &   \ =\   \dot{E}_i - B_i \,,\\ 
\Psi &  \ =\    - C + \frac{1}{3} \Delta E - H ( B -   \dot{E} )  \,,\\
\Phi & \ =\   \phi + H ( B - \dot{E} ) + \dot{B} - \ddot{E}  \,,\\
\Theta &  \ =\   \varphi + \dot{\mathcal{X}}^{(0)}  ( B - \dot{E} )  \,.
\end{split} 
\end{equation}
In contrast to pure gravity on flat space, the gauge invariant scalar modes $\Psi$ and $\Phi$ include Hubble corrections and there is one additional gauge invariant degree of freedom due to the introduction of the scalar matter field. 
\subsection{Homotopy Interpretation}
We now extend the discussion in Section \ref{sectionhomotopyinterpretationlingrav} and provide the homotopy interpretation of gauge invariant variables in cosmological perturbation theory. Let us consider the chain complex,
  \be 
\begin{array}{ccccccccccc} 
X_1 & \xlongrightarrow{\partial} &X_0 {}  
\\[1.5ex]
\{\xi_{\mu}\}& &\{h_{\mu\nu}\;, \; \varphi \} &  
\end{array}
\ee
with an abstract differential $\partial$ mapping the space of gauge parameters to the space of fields. The differential acts as:  
 \be\label{AbstrdifferentialCosmo}
 \begin{split}
  \partial(\xi)_{\mu\nu}&=\partial_{\mu}\xi_{\nu}+\partial_{\nu}\xi_{\mu}-2H \xi_0 \eta_{\mu\nu}\;,\\
  \partial(\xi)_{\bullet} &= -\dot{\mathcal{X}}^{(0)} \xi_0\;,
  \end{split}
 \ee 
 where the $\bullet$ indicates the component in the direction of the scalar field $\varphi$.
In terms of the Bardeen variables in (\ref{BardeenVarsCosmo}), the projected fields in $\bar{X}_0$ are
  \be\label{intermsofBardeenCosmo}
   \bar{h}_{ij}=\widehat{h}_{ij}-2\Psi\delta_{ij}\;, \qquad \bar{h}_{0i}=-\Sigma_i\;, \qquad \bar{h}_{00}=-2\Phi\;,\qquad \bar{\varphi} = \Theta\;.
  \ee
Because of the constraints in (\ref{constraintsonBEandh}), the projected fields identically satisfy: 
\begin{equation}
\label{gaugefixingcosmo}
\partial^i \bar{h}_{ij} - \frac{1}{3} \partial_j ( \delta^{kl}\bar{h}_{kl} ) = 0 \;, \quad \partial^i\bar{h}_{0i} = 0 \,. 
\end{equation}
 Following the same procedure as in Section \ref{sectionhomotopyinterpretationlingrav}, we write the original fields in terms of the projected fields plus pure gauge terms:
 \begin{flalign} \label{hintobarhCosmo}
  h_{\mu\nu}&=\bar{h}_{\mu\nu}+\partial_{\mu}A_{\nu}+\partial_{\nu}A_{\mu}-2 H A_0 \eta_{\mu\nu}\;, \\
  \label{varphi into bar varphi }
  \varphi &= \bar{\varphi} - \dot{\mathcal{X}}^{(0)}A_0 \;, 
  \end{flalign}
where 
 \be\label{AintermsofhCosmo}
  A_{\mu}=(A_0,A_i)=(B- \dot{E}, E_i+ \partial_iE)\;.
 \ee 
In order to find the homotopy map $s: X_0\rightarrow X_1$, we compute 
\begin{flalign}
   (\iota p-{\rm id})(h_{\mu\nu})& = \bar{h}_{\mu\nu}-h_{\mu\nu}=-\partial_{\mu}A_{\nu}-\partial_{\nu}A_{\mu} + 2 H A_0 \eta_{\mu\nu}= -\partial(A)_{\mu\nu}\;, \\
   (\iota p-{\rm id})(\varphi) &= \bar{\varphi}-\varphi=\dot{\mathcal{X}}^{(0)} A_0 =- \partial(A)_\bullet\;,
 \end{flalign} 
from which we can infer:
 \begin{equation}
 \begin{split}
  s(h)_{\mu}&= A_{\mu}\in X_1\,,\\
  s(\varphi)& = A_\bullet\in X_1 \,.
 \end{split}
 \end{equation}
We note that the variable (\ref{AintermsofhCosmo}) has also appeared in eq.~(50) in  \cite{Abramo:1997hu}.

\subsection{Quadratic action in terms of Bardeen variables} 
In order to compute the quadratic action around FLRW we can use the vielbein formalism. We introduce the vielbein, which satisfies 
\begin{equation}
e_\mu{}^a e_{\nu}{}^{b} \eta_{ab} = g_{\mu\nu} \;,
\end{equation}
and its inverse $e_{a}{}^\mu$ is defined by $e_{a}{}^\mu e_{\mu}{}^b = \delta_a{}^b$ and $e_{\mu}{}^a e_{a}{}^\nu = \delta_{\mu}{}^\nu$. 
The Einstein-Hilbert action can be expressed in terms of the vielbein as:
  \be
  \label{quadratic action with Omegas}
 \int d^4x\  \sqrt{-g}\ R \ =\  \int d^4x \,e\Big(-\frac{1}{4}\Omega^{abc}\Omega_{abc}+\frac{1}{2}\Omega^{abc}\Omega_{bca}+\Omega_a\Omega^a\Big)\;, 
  \ee
where $e$ is the determinant of the vielbein and
 \be\label{anholonomycoeff}
  \Omega_{abc}\ \equiv \ e_{a}{}^{\mu} e_{b}{}^{\nu}(\partial_{\mu}e_{\nu c}-\partial_{\nu}e_{\mu c})\,, \qquad \Omega_a \ \equiv \  \Omega_{ab}{}^{b}\,
 \ee
 are the anholonomy coefficients. The vielbein is expanded as
 \be
  e_{\mu}{}^{a}(\eta,{\bf x})\ =\ \bar{e}_{\mu}{}^{a}(\eta) +a(\eta) h_{\mu}{}^{a}(\eta, {\bf x})\;, 
 \ee 
where 
 \be
 \bar{e}_{\mu}{}^{a} (\eta)\ =\  a(\eta) \begin{pmatrix}
1& 0 \\
 0 & \delta_{i}{}^\alpha
 \end{pmatrix} 
 \ee
is the background FLRW frame which satisfies $\bar{e}_{\mu}{}^a \bar{e}_{\nu}{}^b \eta_{ab} =   \bar{g}_{\mu\nu}$,
 and 
 \be
 {h}_{\mu}{}^{a} \ = \ \begin{pmatrix} h_{0}{}^{\bar{0}} & h_{0}{}^{\alpha} \\ h_{i}{}^{\bar{0}} & h_{i}{}^{\alpha} \end{pmatrix}
 \ = \ \begin{pmatrix} \phi &\mathcal{B}^{\alpha} \\ 0 & h_{i}{}^{\alpha} \end{pmatrix}
 \ee
is the (rescaled) fluctuation.  Here we performed a $3+1$ split of indices:
 \be
   \mu \ =\ (0,i)\,, \quad a\ =\ (\bar{0},\alpha)\,, 
 \ee
and picked a gauge for the local Lorentz transformations with  $h_{i}{}^{\bar{0}}=0$. 
The computation of the quadratic action requires up to second order in fluctuations of the inverse vielbein:
 \be
  e_{a}{}^{\mu} \ =\  \bar{e}_{a}{}^{\mu}-a(\eta) \bar{e}_{a}{}^{\nu} h_{\nu}{}^{b} \bar{e}_{b}{}^{\mu} + a^2(\eta) \bar{e}_{a}{}^\nu h_{\nu}{}^b \bar{e}_{b}{}^\rho h_{\rho}{}^c \bar{e}_{c}{}^\mu\,. 
 \ee
Writing this in components we can summarize the vielbein and its inverse as 
 \be
 \begin{split}
   {e}_{\mu}{}^{a} &\ =\  \begin{pmatrix} e_{0}{}^{\bar{0}} & e_{0}{}^{\alpha} \\ e_{i}{}^{\bar{0}} & e_{i}{}^{\alpha} \end{pmatrix}
\  =\   a(\eta) \begin{pmatrix}  1+ \phi  &\mathcal{B}^{\alpha} \\ 0 & \delta_{i}{}^{\alpha}+ h_{i}{}^{\alpha} \end{pmatrix}\;, \\
 e_{a}{}^{\mu} &\ =\   \begin{pmatrix} e_{\bar{0}}{}^{{0}} & e_{\bar{0}}{}^{i} \\ e_{\alpha}{}^{{0}} & e_{\alpha}{}^{i} \end{pmatrix}
 \ =\   a^{-1}(\eta) \begin{pmatrix}  1- \phi + \phi^2 & -\mathcal{B}^{i} + \phi \mathcal{B}^i + \mathcal{B}^j h_{j}{}^\alpha \delta_{\alpha}{}^i  \\ 0 & \delta_{\alpha}{}^{i} - h_{\alpha}{}^{i} + h_{\alpha}{}^j h_{j}{}^\beta \delta_{\beta}{}^i  \end{pmatrix} \,. 
 \end{split}
 \ee
This vielbein leads to the same parameterizations of the first order fluctuations of the metric that is standard in cosmology, namely in (\ref{SVTdecompositionTensor}), 
with $h^{0i}$ identified with $\mathcal{B}^{\alpha}=-h^{0i}\delta_{i}{}^{\alpha}$ via the background vielbein, and with  
 \be
  h_{ij} \ =\ 2\, h_{(i}{}^{\alpha}\delta_{j)\alpha}\,.
 \ee
To second order, we collect the fluctuations of the metric and its inverse:
\begin{equation}
\begin{split}
g_{00}& \ =\  a^2 ( -1 -2 \phi - \phi^2 + \mathcal{B}^\alpha\mathcal{B}_\alpha)\,,\\
g_{0i}&\  =\  a^2 ( \mathcal{B}_i + \mathcal{B}_{\alpha} h_{i}{}^\alpha)\,,\\
g_{ij} &\ = \ a^2 ( \delta_{ij} + h_{ij} + h_i{}^\alpha h_{j\alpha})\,, \\
g^{00}& \ =\  a^{-2} ( -1 +2 \phi - 3\phi^2  )\,,\\
g^{0i}& \ = \ a^{-2}( \mathcal{B}^i -2\phi \mathcal{B}^i - \mathcal{B}^j h_j{}^\alpha \delta_\alpha{}^i) \,,\\ 
g^{ij} &\ = \ a^{-2}  (\delta^{ij}- 2 h_{\alpha}{}^{(i} \delta^{|\alpha|j)}+ 2 h_{\alpha}{}^k h_{k}{}^\gamma \delta_\gamma{}^{(i} \delta^{|\alpha|j)})\,.
\end{split}
\end{equation}
The SVT decomposition in (\ref{SVTdecompositionTensor})  translates to that of $h_{i}{}^\alpha$ as
 \be
  h_{i}{}^{\alpha} \ = \ \widehat{h}_{i}{}^{\alpha} + \partial_i E^{\alpha}+\partial_i\partial^{\alpha} E+\delta_{i}{}^{\alpha}\Big(C-\frac{1}{3}\Delta E\Big)\,,
 \ee
with $\widehat{h}_{i}{}^\alpha$ satisfying the constraints:
\begin{equation}
 \delta_{\alpha}{}^i \widehat{h}_{i}{}^{\alpha} \ =\  0 \,,\quad \partial^i \widehat{h}_{i}{}^{\alpha} \ =\  0  \,.
\end{equation}
Similarly, the vector $\mathcal{B}^{\alpha}$ is decomposed as:
\begin{equation}
\mathcal{B}^\alpha \ =\  B^\alpha + \partial^\alpha B \,, \quad \partial_\alpha B^\alpha \ =\  0 \,. 
\end{equation}
Inserting the decompositions into (\ref{quadratic action with Omegas}), we compute the full action with matter coupling (\ref{Full matter action cosmology}) to quadratic order. The linear terms of the action drop out, assuming the background field equations are satisfied. The quadratic action is:
 \begin{equation}
 \begin{split}
 \label{raw quadratic action}
 S & = \int d^4 x\; a^2 \bigg\{ 
 \frac{1}{4} \dot{h}^{ij} \dot{h}_{ij} 
 +   \frac{1}{4} h_{ij}\Delta h^{ij}
  +\frac{1}{2}  \partial_j h^{ij}    \partial^k h_{ik}  + \frac{1}{2} ( h_{i}{}^i -h_{00}) \partial^j \partial^k h_{jk}
  \\
  & 
\quad \quad \quad \quad \quad \;\;+ \frac{1}{2} h_{0i}\Delta h^{0i}  -\frac{1}{2} ( \partial_i h^{0i})^2  
 + \partial^i h^{0j} \dot{h}_{ij}
 +  \partial^i h_{0i} \dot{h}_{j}{}^j  
 \\
 &\quad \quad \quad \quad \quad \;\;- \frac{1}{4} (\dot{h}_{i}{}^{i })^2  -  \frac{1}{4} h_{i}{}^i \Delta   h_{j}{}^j 
 +\frac{1}{2}  h_{i}{}^i \Delta h_{00}   
   \\
 &    \quad \quad \quad \quad \quad \;\;    -\frac{1}{2} ( \dot{H} +2H^2) h_{00}^2      
   - H h_{00} \dot{h}_{i}{}^i 
+2 H h_{00} \partial^ih_{0i}
\\
  &\quad \quad \quad \quad \quad \;\;+ \frac{1}{2} \dot{\varphi}^2+ \frac{1}{2}  \varphi  \Delta  \varphi- \frac{1}{2} a^2V''({\mathcal{X}}^{(0)}) \varphi^2 
\\&
 \quad \quad \quad \quad \quad \;\;   -\frac{1}{2} \dot{\mathcal{X}}^{(0)}\varphi \big(  \dot{h}_{00} + \dot{h}_{i}{}^i +2 \partial_i h^{0i}  \big) +a^2 V'({\mathcal{X}}^{(0)}) \varphi h_{00}
 \bigg\} \;.
 \end{split}
 \end{equation}
  
By inserting the Bardeen variables in (\ref{BardeenVarsCosmo}), after a tedious computation the quadratic action can be organized into its gauge invariant form:
\begin{equation}
\begin{split}
\label{GIquadraticCosmo}
S&\ =\ \int d^4 x \;  a^2 \; \bigg\{ \frac{1}{4}
  \dot{\widehat{h}}^{ ij }\dot{\widehat{h}}_{ ij }   
   +  \frac{1}{4} \widehat{h}^{ ij }\Delta  \widehat{h}_{ij}   - \frac{1}{2} \Sigma_i\Delta \Sigma^i
 + 4 \Psi \Delta \Phi  -2 \Psi \Delta \Psi  + \frac{1}{2} \dot{\mathcal{X}}^{(0)2} \Phi^2   -6 ( \dot{\Psi} + H \Phi)^2 \\&\quad \quad \quad \quad \quad \quad \ \
  +  \frac{1}{2} \dot{\Theta}^2 + \frac{1}{2} \Theta \Delta \Theta- \frac{1}{2} a^2V''( {\mathcal{X}}^{(0)}) \Theta^2 + \dot{\mathcal{X}}^{(0)}\Theta ( \dot{\Phi } + 3 \dot{\Psi} ) - 2 a^2 V'( {\mathcal{X}}^{(0)}) \Theta \Phi 
   \bigg\}\;. 
 \end{split}
\end{equation}

The lengthy computation of organizing (\ref{raw quadratic action}) into gauge invariant form is dramatically shortened by substituting the fields with their projections plus pure gauge terms, (\ref{hintobarhCosmo}) and (\ref{varphi into bar varphi }), into the action. All the pure gauge terms drop out by  gauge invariance. Then by inserting the variables  (\ref{intermsofBardeenCosmo}), one quickly arrives at the manifestly gauge invariant action (\ref{GIquadraticCosmo}). 
As a further consistency check (beyond the fields combining into gauge invariant objects) we set to zero all modes but the scalar modes 
in order to compare with the results of \cite{Mukhanov:1990me}.
Deconstructing the scalar Bardeen variables in (\ref{GIquadraticCosmo}), i.e.~by expressing them in terms of the scalars ($C$, $E$, $B$, $\phi$, and $\varphi$) as in (\ref{BardeenVarsCosmo}), we indeed reproduce the quadratic action  (10.68) obtained in \cite{Mukhanov:1990me}.

Finally, we use the above result in order to re-derive the Mukhanov-Sasaki action that governs the scalar modes in a gauge invariant manner. 
We will thus set in the action all modes to zero except for the scalar modes. It is moreover convenient to introduce the following combination 
of gauge invariant variables: 
\begin{equation}
W \equiv \Theta +  f \Psi\,, \quad   \text{where} \quad f \equiv \frac{\dot{\mathcal{X}}^{(0)}}{H}\,.
\end{equation}
Substituting $\Theta$ in terms of $W$ and $\Psi$, the action becomes:
\begin{equation}
\begin{split}
S = \int d^4 x \;  a^2 \; \bigg\{&
   \frac{1}{2}\dot{W}^2 + \frac{1}{2} W \Delta W- \frac{1}{2} a^2 V'' W^2\\& + 4 \Psi \Delta \Phi  -2 \Psi \Delta \Psi- f \Psi \Delta W  + \frac{1}{2}f^2 \Psi \Delta \Psi
  \\&
+ \frac{1}{2}f^2\bigg( -\dot{H} + 2 H^2 + \frac{2 \ddot{H}}{H}- \frac{2\dot{H}^2}{H^2} + \frac{6\ddot{\mathcal{X}}}{f} \bigg)\Psi^2   \\&  
 - \dot{W} ( \dot{f} \Psi + f \dot{\Psi} )  + a^2 V'' f \Psi W + 3 \dot{\mathcal{X}} W \dot{\Psi} \\
 & 
   + \bigg(\frac{1}{2}f^2  -6\bigg) (\dot{\Psi}+ H \Phi )^2
-f^2 (\dot{H} + 2 H^2)
  \Psi \Phi   
 + \dot{\mathcal{X}} W  \dot{\Phi } - 2 a^2 V'  W \Phi 
   \bigg\} \,. 
 \end{split}
\end{equation}
We have omitted the superscript on the background quantity $\mathcal{X}^{(0)}$ for convenience in notation. 
It now turns out that the dependence on $\Phi$ can be removed by performing the field redefinition 
\begin{equation}
\Phi\rightarrow \Phi = \Upsilon - \frac{1}{H} \dot{\Psi} + \frac{f}{4} W - \frac{f^2}{4} \Psi \,. 
\end{equation}
The action then reduces to a function of only $\Psi$, $\Upsilon$, and $W$:
\begin{equation}
\begin{split}
\label{almost MS}
S= \int d^4 x \;  a^2 \; \bigg\{&
    \frac{1}{2}\dot{W}^2 + \frac{1}{2} W \Delta W+ \frac{1}{2} \bigg(- a^2 V'' -2 \dot{H} + 4 H^2 + \frac{2\dot{H}^2}{H^2}- \frac{2\ddot{H}}{H}\bigg)W^2\\& + 4 \Psi \Delta \Upsilon +\bigg(\frac{1}{2}f^2  -6\bigg) H^2 \Upsilon^2 
 + \bigg( \ddot{\mathcal{X}} - \frac{\dot{H}}{H} \dot{\mathcal{X}}
  \bigg) \Upsilon W  - \dot{\mathcal{X}} \Upsilon \dot{W}  
   \bigg\} \,. 
 \end{split}
\end{equation} 
The second line of (\ref{almost MS}) can be simplified by replacing
\begin{equation}
\Psi\rightarrow \Psi = \Gamma
-  \frac{1}{4}\Delta^{-1} \bigg[
 \bigg(\frac{1}{2}f^2  -6\bigg) H^2 \Upsilon 
+ \bigg( \ddot{\mathcal{X}} - \frac{\dot{H}}{H} \dot{\mathcal{X}} \bigg) W - \dot{\mathcal{X}} \dot{W}  \bigg] \,, 
\end{equation} 
so that 
\begin{equation}
\begin{split}
\label{nearlyMS}
S = \int d^4 x \;  a^2 \; \bigg\{&
    \frac{1}{2}\dot{W}^2 + \frac{1}{2} W \Delta W+ \frac{1}{2} \bigg(- a^2 V'' -2 \dot{H} + 4 H^2 + \frac{2\dot{H}^2}{H^2}- \frac{2\ddot{H}}{H}\bigg)W^2  + 4 \Gamma \Delta \Upsilon
   \bigg\}\,.
\end{split}
\end{equation}
We next introduce   the Mukhanov variable $v\equiv a W$ and bring the last term in (\ref{nearlyMS}) to a  diagonal form by defining 
\begin{equation}
\Phi_{\pm}\equiv a ( \Gamma \pm \Upsilon) \,,
\end{equation}
to obtain
\begin{equation}
\begin{split}
S  = \int d^4 x \;   \; \bigg\{&
 \frac{1}{2} v\Delta v + \frac{1}{2} \dot{v}^2 + \frac{1}{2} \frac{\ddot{z}}{z} v^2  + \Phi_+ \Delta \Phi_+ - \Phi_- \Delta \Phi_- 
   \bigg\} \,,\quad \text{where} \quad z\equiv a f \,. 
\end{split}
\end{equation} 
Finally, by using the invertibility of the Laplacian, the equations of motion for $\Phi_+$ and $\Phi_-$ give $\Phi_+=\Phi_-=0$, 
and re-substituting into the action we obtain 
\begin{equation}
\label{Mukhanov Sasaki}
S = \int d^4 x \;   \; \bigg\{ \frac{1}{2} v \Delta v + \frac{1}{2} \dot{v}^2 + \frac{1}{2} \frac{\ddot{z}}{z} v^2 \bigg\} 
\,. 
\end{equation}
This is the Mukhanov-Sasaki action for the gauge invariant scalar mode.

\section{Discussion and Outlook}

The main result of this paper is to provide a completely systematic approach towards formulating gauge theories perturbatively in terms of gauge invariant variables 
to arbitrary order. We have shown that passing over to gauge invariant variables, such as those used in cosmological perturbation theory,  can be interpreted 
in terms of homotopy transfers 
in the framework of the $L_{\infty}$ formulation of gauge theories. The perturbation lemma then provides an 
algorithmic procedure to determine gauge invariant variables to arbitrary order. 
We have also explained how gauge invariance allows one to write the action in terms of gauge invariant variables by replacing the 
original fields by gauge invariant ones. 
Since the gauge invariant variables satisfy constraints formally identical 
to familiar gauge fixing conditions (Coulomb gauge for Yang-Mills theory and transverse gauge for gravity), our results provide an effective procedure 
to take an action determined in those gauges and recover a fully gauge invariant action expressed in terms of gauge invariant variables.

We will now  argue more generally that working in a particular  gauge  can  be reinterpreted in terms of suitable gauge invariant 
variables that are non-local functions of the original fields. 
(See also \cite{Barnich:2005ga} for how to pass over to field variables in light-cone gauge for the BV-BRST formulation of free higher-spin theories.) 
Consider, for instance, linearized gravity 
in \textit{synchronous gauge}, which amounts to setting  $h_{00}=h_{0i}=0$. 
Can one reinterpret the resulting action in gauge invariant terms? This is possible, but only at the cost of 
introducing a  non-locality  in time. 
Given an arbitrary $h_{\mu\nu}$, subject to all gauge redundancies,  we define 
 \be
  A_0(t, x) := \frac{1}{2}\int^t_{t_0} dt'\,h_{00}(t',x)\; , 
 \ee
where $t_0$ is an arbitrary reference time. 
Under linearized diffeomorphisms  (\ref{linDIff}) we then have  
 \be
  \delta_{\xi}A_0=\int^t_{t_0} dt'\,\partial_{t'}\xi_0(t') = \xi_0\,,
 \ee
where we set $\xi_{\mu}(t_0)=0$. We can next  define 
 \be
  A_i(t,x):=\int_{t_0}^t dt'(h_{0i}(t',x)-\partial_i A_0(t',x))\,, 
 \ee
which transforms as $\delta_{\xi}A_i=\xi_i$. In total, we then have $\delta_{\xi}A_{\mu}=\xi_{\mu}$, 
and so gauge invariant variables are given by 
 \be\label{gaugeinvsynchron}
 \bar{h}_{\mu\nu}\equiv h_{\mu\nu}-\partial_{\mu}A_{\nu}-\partial_{\nu}A_{\mu}\,, 
 \ee
where the $A_{\mu}$ are the above non-local functions of $h_{\mu\nu}$.  
It follows from this definition that the gauge invariant variables satisfy 
$\bar{h}_{00}=0$ and $\bar{h}_{0i}=0$, exactly imitating  the synchronous gauge. 
Conversely, from the action in synchronous gauge one may reconstruct 
the fully gauge invariant action by reinterpreting the field as the right-hand side of (\ref{gaugeinvsynchron}). 
We may again interpret the passing over to gauge invariant variables as a homotopy transfer, 
with the homotopy map $s(h)_{\mu}=A_{\mu}$. 
We learn that  gauge invariant variables exist in various guises. 
In cosmology, however,  the Bardeen variables seem to be preferred because they respect the symmetries of the 
FLRW backgrounds.

While our  discussion shows that there is an operational equivalence between gauge fixing and finding gauge invariant variables, 
we hope that the systematic approach introduced in this paper may help alleviate some of the interpretational issues 
that seem to have arisen particularly in cosmological perturbation theory.  More generally, we expect that the general framework 
of $L_{\infty}$ algebras and homotopy transfer may become useful in setting up the perturbation theory of various problems 
in gravitational physics, for instance, in computing higher-point cosmological correlation functions in a manifestly gauge invariant manner. 
We hope to report on such results in the not too distant future.

\subsection*{Acknowledgements}

This work is funded  by the ERC Consolidator Grant ``Symmetries \& Cosmology" 
and by the Deutsche Forschungsgemeinschaft (DFG, German Research Foundation) - Projektnummer 417533893/GRK2575 ``Rethinking Quantum Field Theory".

\end{document}